\documentclass[reprint,superscriptaddress,aps,prl,twocolumn]{revtex4-2}
\usepackage[T1]{fontenc}
\usepackage{graphicx}
\usepackage{array}
\usepackage{makecell} 
\usepackage{tabularray}
\usepackage{booktabs}
\usepackage{dcolumn}
\usepackage{amsmath}
\usepackage{amssymb}
\usepackage{amsfonts}
\usepackage{bm}
\usepackage{mathtools} 
\usepackage{mathrsfs}
\usepackage{bbm}
\usepackage[normalem]{ulem}
\usepackage{physics}
\usepackage[dvipsnames,x11names]{xcolor}
\usepackage{csquotes}
\usepackage[colorlinks=true,allcolors=blue]{hyperref}
\usepackage{notes2bib}
\bibnotesetup{note-name = , use-sort-key = false}

\usepackage{orcidlink}

\newcommand{\sectitle}[1]{\textit{#1}---}
\newcommand{\fref}[2][]{\hyperref[#2]{\ref*{#2}#1}}
\newcommand{\dbar}{{\mathchar'26\mkern-10mu\delta}}
\newcommand{\ddbar}{\mathrm{d}{\mkern-7mu\mathchar'26\mkern-2mu}}

\newcommand{\qv}{{\vb{q}}}

\newcommand{\xv}{{\vb{x}}}
\newcommand{\yv}{{\vb{y}}}

\newcommand{\mob}{M}
\newcommand{\symm}{\mathsf{S}}

\newcommand{\connected}{\mathrm{c}}

\newcommand{\Mc}{\hat M}
\newcommand{\Lc}{\hat L}
\newcommand{\Kc}{\mathcal{K}}
\newcommand{\Sc}{\mathcal{S}}

\newcommand{\Fc}{\mathcal{F}}

\newcommand{\Qv}{\vb{Q}}
\newcommand{\Qvs}[1]{\vb{q}^\Sigma_{#1}}

\newcommand{\Yv}{\vb{Y}}
\newcommand{\Uc}{\mathcal{U}}
\newcommand{\lin}{L}
\newcommand{\G}{u}
\newcommand{\hu}{\hat{u}}

\newcommand{\N}{N}

\newcommand{\aA}{a_\mathrm{A}}
\newcommand{\aB}{a_\mathrm{B}}
\newcommand{\nonlin}{{m}}
\newcommand{\mrn}{{\mathfrak{m}(r,n)}}
\newcommand{\noise}{\epsilon}
\newcommand{\NRCHone}{\varphi}
\newcommand{\NRCHtwo}{\psi}
\newcommand{\density}{\varphi}
\newcommand{\pol}{p}
\newcommand{\vel}{v}
\newcommand{\prhoc}{z}
\NewDocumentCommand{\ddq}{o}{%
\ddbar\qv\IfValueT{#1}{_{#1}}\,
}%

\newcommand{\pddqphi}[1]{\prod_{i=1}^{#1}[\ddq[i]\phi_{\qv_i}]\,}
\newcommand{\brq}[1]{\bigl(\Qv_{#1}\bigr)}
\newcommand{\firstSM}{\bibnote[SM]{See Supplemental Material below.}}
\newcommand{\SM}{\bibnotemark[SM]}

\begin{document}

\clearpage
\title{Generic nonlocal statistics of the stationary measure in conserved active systems}

\author{Filippo De Luca\,\orcidlink{0000-0003-3849-343X}
}
\affiliation{DAMTP, Centre for Mathematical Sciences, University of Cambridge, Wilberforce Road, Cambridge CB3 0WA, United Kingdom}

\author{Michael E. Cates\,\orcidlink{0000-0002-5922-7731}
}
\affiliation{DAMTP, Centre for Mathematical Sciences, University of Cambridge, Wilberforce Road, Cambridge CB3 0WA, United Kingdom}

\author{Cesare Nardini\,\orcidlink{0000-0002-0466-1418}
}
\affiliation{Service de Physique de l'\'Etat Condens\'e, CEA, CNRS Universit\'e Paris-Saclay, CEA-Saclay, 91191 Gif-sur-Yvette, France}
\affiliation{Sorbonne Universit\'e, CNRS, Laboratoire de Physique Th\'eorique de la Mati\`ere Condens\'ee, 75005 Paris, France}

\begin{abstract}
The stationary measure of equilibrium systems with detailed balance follows a Boltzmann distribution, so that for short-ranged interactions the measure is local, meaning that distant spatial domains are statistically independent. In contrast, active systems break detailed balance, and can have nonlocal stationary measure even for fully local dynamics. Here, by expanding in nonlinearity about a Gaussian-model limit, we construct the measure perturbatively deep in the disordered phase for a class of models that includes Active Model A, Active Model B+, Model AB, the Nonreciprocal Cahn--Hilliard model, and the Toner--Tu model. In this regime, nonlocality is linked to a dynamical conservation law. Our results generically preclude construction of a Landau--Ginzburg expansion of the stationary measure (as opposed to the dynamical equations) for conserved active field theories.

\end{abstract}
\maketitle

\sectitle{Introduction.}Active matter evades the constraints of thermal equilibrium through dissipative energy fluxes at the microscopic scale~\cite{ramaswamy-2010,marchetti-2013,gompper-2020,vrugt-2025}. Active systems can therefore display a wide range of phenomena disallowed in equilibrium. For example, in self-propelled colloids~\cite{palacci-2013,vanderlinden-2019}, biomolecular condensates~\cite{hyman2014liquid,banani2017,berry2018physical,Weber2019review}, and cell-sorting tissues~\cite{balasubramaniam-2021}, bulk phase separation can arrest to give micro\-phase separation, even in the absence of the long-range interactions that cause this in equilibrium~\cite{cates-2025}. Flocking models, addressing herds of wildebeest and starling flocks, can exhibit long-range order even in two dimensions, where the Mermin--Wagner theorem would forbid it in equilibrium~\cite{toner-1998}. Non-reciprocal systems, such as living crystals~\cite{tan-2022} and asymmetrically interacting binary mixtures~\cite{meredith-2020,dinelli-2023,alharraq-2025}, show statistically stationary states involving collective oscillations and travelling waves.

An important toolbox comprises active field theories, which address collective degrees of freedom, such as the density of constituent agents and/or their average polarization. These generic descriptions, which often expand equations of motion at Landau--Ginzburg level, can capture the emerging physics due to activity, independently of microscopic details. They have explained theoretically many of the phenomena just mentioned, including microphase separation~\cite{tjhung-2018,fausti-2021}, capillary instabilities~\cite{fausti-2021}, the flocking transition~\cite{toner-1998}, and non-reciprocal steady states~\cite{you-2020,saha-2020,brauns-2024}. Furthermore, they have allowed to identify hallmarks of activity, such as entropy production~\cite{nardini-2017a}, steady-state probability currents~\cite{obyrne-2023}, and time-reversal symmetry breaking~\cite{obyrne-2025a}.

In this Letter, we address another key signature of activity, namely the nonlocality of the steady-state probability distribution $P_\infty$, or equivalently its logarithm (which can be viewed as an effective free energy functional for static purposes)
$\Uc\coloneqq -\noise\ln P_\infty$, where $\noise$ is the noise strength. Its small-noise limit $\Uc_0$ is the quasipotential of Large Deviation Theory (LDT)~\cite{freidlin-2012}, a connection we exploit below. We define $P_\infty$ to be \emph{local} if different spatial regions are connected in $\Uc$ by a kernel of finite range, so that distant domains are statistically independent (modulo global conservation laws) and \emph{nonlocal} if $\Uc$ has a kernel with power-law decay. In thermal (Boltzmann) equilibrium, if the Hamiltonian is short-ranged, $P_\infty$ is local, but out of equilibrium this is no longer true. In a few instances of nonequilibrium models where analytical calculation of $\Uc$ was possible, either via an equilibrium mapping~\cite{clincy-2003,ayyer-2009,li-2020b}, or in the weak-noise, large-deviation limit~\cite{derrida-2002,enaud-2004,bertini-2005,bertini-2015,woillez-2020}, $P_\infty$ was indeed found to be nonlocal.

\begin{figure}[tbp]
\centering
    \includegraphics[width=\columnwidth]{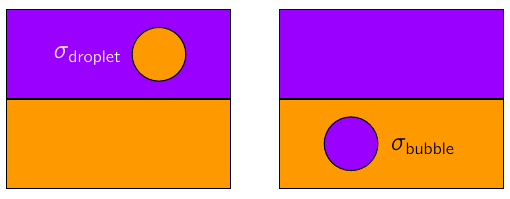}
    \caption{\label{fig:lowtemp}In AMB+ at bulk phase coexistence (vapor phase purple, liquid yellow), liquid-droplet in vapor (left) and vapor-bubble in liquid (right) excitations are controlled by different interfacial tensions $\sigma_\mathrm{droplet},\,\sigma_\mathrm{bubble}$~\cite{cates-2023}. This requires a nonlocal stationary measure in the low-temperature phase~\cite{gibbs-1876,gibbs-1878,tolman-1948}. However, finding the precise character of this nonlocality is intractable.}
\end{figure}

Likewise, for certain active field theories in specific regimes, nonlocality
of $P_\infty$ can be established nonperturbatively. For example, consider the regime of bulk phase separation in Active Model B+ (AMB+, an extension of $\phi^4$ dynamical field theory). Here, if $\Uc[\phi]$ were local, in studying phase separation it would play a role identical to the Helmholtz free energy functional of an equilibrium fluid. It has been known since Gibbs \cite{gibbs-1876,gibbs-1878,tolman-1948} that for a fluid with short-range interactions, and thence local measure, this can be decomposed into a bulk part $\Uc_\mathrm{bulk}$ (which is equal in the two phases at liquid-vapor coexistence) and an interface part $\Uc_\mathrm{int} = \sigma A$, where $A$ is the interfacial area and $\sigma$ the interfacial tension.
However, a low-noise study of AMB+ shows that the quasipotentials $\Uc_0$ of liquid droplets in vapor, and vapor bubbles in liquid, are governed by two different tensions $\sigma_\mathrm{droplet},\,\sigma_\mathrm{bubble}$~\cite{cates-2023}. The stationary measure at coexistence thus assigns different probabilities to droplet and bubble excitations of equal radius (see Fig.~\ref{fig:lowtemp}); this, via Gibbs, requires nonlocal $\Uc$. 

Separately, recent numerical studies found a nonlocal $P_\infty$ in the {\em microphase-separated} regime of AMB+~\cite{brossollet-2025a} (but not so far the bulk phase separation regime just considered). A full characterization of the nonlocality in either of these strongly nonlinear regimes appears intractable, and more generally a
systematic framework to address the emergence of nonlocality in active systems remains lacking. We achieve this below, albeit perturbatively and for the non-separated, disordered phases of such models.

Another signature of nonlocality in $P_\infty$ is the emergence of long-range correlations unlinked to phase transitions. In equilibrium systems with local Gibbs--Boltz\-mann measure, power-law correlators appear generically at critical points and, if a continuous symmetry is present, across the ordered phase for the transverse directions of the order parameter~\cite{forster-2018}. Out of equilibrium, long-range correlators emerge far more broadly: in the ordered phase of active polar and nematic fluids, a fact related to so-called giant number fluctuations~\cite{ramaswamy-2003,toner-2005,chate-2008,cavagna-2010,cavagna-2015,ginelli-2016}; in two-temperature systems \cite{damman-2024};
in systems driven at the boundary~\cite{procaccia-1979,spohn-1983,bertini-2009}; and in active systems with quenched disorder~\cite{ro-2021}. Importantly, non-critical power laws were found to emerge {\em generically} in non-equilibrium systems with a conservation law~\cite{grinstein-1990,garrido-1990,grinstein-1991,grinstein-1993,dorfman-1994}.

However, the implications of those findings for the stationary measure have to date remained unexplored. One reason for this is that computing $P_\infty$ far from equilibrium is generically an impossible task. Here, we tackle this issue by expanding (in the high-temperature phase) about the Gaussian limit of generic active field theories, where both the noise and the active and passive nonlinearities are small. There, we employ and extend a perturbative scheme rooted in LDT~\cite{freidlin-2012,hugang-1988,tel-1989,bouchet-2016} to compute the stationary measure order by order in the noise and nonlinearity. 

We show that in the presence of a conservation law, $\Uc$ becomes nonlocal; in contrast, $\Uc$ is short-ranged (in the given perturbative limit) in the absence of a conservation law, or when detailed balance is restored. We further show how a nonlocal $P_\infty$ is reflected in long-range $\N$-point functions. We apply our results to well-studied active models for a single scalar field, including Active Model A, AMB(+), and Model AB. Finally, we turn to theories with more than one order parameter, showing that our results stay valid beyond the single-field case. Concretely, we consider the nonreciprocal Cahn--Hilliard (NRCH) model and Toner--Tu theory, showing that 
one conserved field is enough to make $P_\infty$ nonlocal for all fields that couple to it.

An important outcome of our work is that the stationary measure of a conserved active field does not in general admit a Landau--Ginzburg expansion. (This is despite using such an expansion in the equations of motion, which all the aforementioned active theories do.) This result, albeit for the high-temperature phase, suggests that any static Landau--Ginzburg approach to criticality in active matter requires careful case-by-case justification~\cite{huang-2023}.

\sectitle{Perturbative calculation.}We next calculate the stationary measure of active field theories for the low-noise, weakly nonlinear regime of interest. We start with a single order parameter that obeys
\begin{equation}
    \partial_t\phi(\xv,t) = -\mob\lin\,\phi + \alpha\Kc[\phi] + \sqrt{2\epsilon}\,\eta\,,\label{eq:dynamics}
\end{equation}
where $\mob$ and $\lin$ are linear differential operators; $\eta$ is a Gaussian zero-mean white noise with variance $\expval{\eta(\xv,t)\eta(\xv',t')} = \mob\delta(\xv-\xv')\delta(t-t')$; and $\Kc$ is a nonlinearity governed by a small activity parameter $\alpha$. While our results generically apply to any dynamics in the form of \eqref{eq:dynamics}, we will focus mainly on the specific active field theories mentioned previously. For convenience, we express the nonlinearity $\Kc$ (containing $g\geq 0$ gradients and $m\geq 2$ powers of the field) in Fourier as
\begin{equation}\label{eq:K}
 \Kc_\qv[\phi]\!=\!\int\pddqphi{\nonlin}\hat K\brq{\nonlin}\,\dbar_{\qv-\Qvs{\nonlin}}\,,
\end{equation}
where $\dbar_\qv = (2\pi)^d\delta(\qv)$, $\ddq=\dd{\qv}/(2\pi)^d$, $\Qv_{k}=(\qv_1,\ldots,\qv_{k})$, $\Qvs{k} \coloneqq \sum_{i=1}^k\qv_i$, and $\hat K$~is a polynomial of order $g$ in the $\qv_i$.

The stationary measure $P_\infty$ at $\alpha=0$ is Gaussian and corresponds to a quadratic free energy $\Uc = (1/2)\int \dd{\xv}\phi \lin\phi:=\Uc_\mathrm{G}[\phi]$. Switching on $\alpha$, detailed balance is broken and $P_\infty$ cannot be computed exactly. This happens even in the LDT limit, where the quasipotential $\Uc_0=\lim_{\epsilon\to0} \Uc$ is known exactly in only a few cases~\cite{kipnis-1982,derrida-2002,derrida-2002a,enaud-2004,bertini-2005,cohen-2012,garrido-2021}.
However, we can find the stationary measure $P_\infty$ (equivalently, $\Uc$) perturbatively when activity $\alpha$ and noise $\epsilon$ are {\em both} small. 

Our analysis generalizes the technique of~\cite{bouchet-2016} for computing the quasipotential perturbatively to any order in $\alpha$ in the small noise limit, $\epsilon\to 0$. Expanding $\Uc$ via the WKB ansatz
\begin{equation}\label{eq:WKB}
\Uc[\phi] = \sum_{r=0}^\infty\sum_{n=0}^\infty \noise^r\alpha^n\,\Uc^{(n)}_r[\phi]\,,
\end{equation}
and inserting this into the stationary Fokker--Planck associated with \eqref{eq:dynamics}, we find a hierarchy of equations for the $\Uc_r^{(n)}$~\firstSM. Each can be solved with the method of characteristics, yielding
\begin{equation}\label{eq:Uc_S}
\Uc_r^{(n)}[\phi] = \int_{-\infty}^0\dd{t}\Sc_r^{(n)}[\Phi^{(0)}(t)]\,,
\end{equation}
where $\Sc_r^{(n)}$ are explicit functionals given in the End Matter, that depend on the $\Uc_r^{(n)}$ from previous orders, and $\Phi^{(0)}(t)$ is the instanton of the unperturbed dynamics, connecting its stable fixed point to the field configuration $\phi(\xv)$.

Equation~\eqref{eq:Uc_S} admits analytic iterative construction of $\Uc$ whenever $\Phi^{(0)}(t)$ is explicitly known, as holds when the unperturbed dynamics is linear, which was assumed in \eqref{eq:dynamics}. Notice that the same perturbative structure emerges even if the perturbation is made close to a nonlinear reference dynamics, but calculation of the $\Uc_r^{(n)}$ is then vastly complicated by the need to first find both its quasipotential $\Uc_0^{(0)}$ and instanton $\Phi^{(0)}(t)$~\cite{bouchet-2016}.

For any system obeying \eqref{eq:dynamics}, the quasipotential at first order in $\alpha$ contains $m+1$ fields and is given by
\begin{subequations}\label{eq:Uc_u}
 \begin{align}
        &\Uc_0^{(1)} = \int\pddqphi{\nonlin + 1}\hu_0^{(1)}\brq{\nonlin + 1}\,\dbar_{\Qvs{\nonlin+1}}\,,\label{eq:Uc}\\
    	&\hu_0^{(1)}\brq{\nonlin+1} =-\frac{\left[\hat{K}\brq{\nonlin}\,\Lc_{\qv_{\nonlin+1}}\right]^\symm}{\sum_{i=1}^{\nonlin + 1} \Mc_{\qv_i} \Lc_{\qv_i}}\,,\label{eq:u}
 \end{align}
\end{subequations}
where $\symm$ indicates full symmetrization with respect to all $\qv_i$, while $\Mc_\qv$ and $\Lc_\qv$ denote the Fourier representations of the linear operators $\mob$ and $\lin$. 

We now make two comments. First, when $\Kc = -\mob\delta{\tilde\Fc}/\delta{\phi}$, so that the full nonlinear model obeys detailed balance, our method returns $\Uc_0^{(1)} = \tilde\Fc$, recovering the correct free energy, $\Fc=\Uc_\mathrm{G}+\alpha\tilde\Fc$ (see End Matter). Second, the method allows us to find higher order corrections $\Uc^{(n)}_r$ for any $r$ and $n$.
Without computing the explicit expressions, our analysis shows that $\Uc_r^{(n)}$ is a homogeneous polynomial in $(\nonlin-1)n + 2(1-r)$ fields~\SM. In general, the full stationary measure $P_\infty$ thus contains an infinite hierarchy of many-body effective interactions.

We next show that the effective free energy $\Uc=-\epsilon\ln P_\infty$ is nonlocal when the noiseless ($\epsilon= 0$) dynamics  in~\eqref{eq:dynamics} conserves the global order parameter ${\phi_0=\int \dd{\xv} \phi(\xv)}$, but that, to any order in perturbation theory, $\Uc$ is local otherwise \footnote{The resummation of the infinite series in \eqref{eq:WKB} is not addressed here.}. We need consider only $\Uc_0^{(1)}$, written in real space as
 \begin{align}\label{eq:Uc_G}
    &\Uc_0^{(1)}\!=\!\int\!\dd{\xv} \phi(\xv) \prod_{i=1}^{\nonlin} [\dd{\yv_i}\phi(\xv+\yv_i)] \G_0^{(1)}(\Yv_m)\,,
 \end{align}
where $\G_0^{(1)}$ is the inverse Fourier transform of $\hu_0^{(1)}$, and $\Yv_{m}=(\yv_1,\ldots,\yv_{m})$. As is standard~\cite{lighthill-1958}, the large-scale properties of $\G_0^{(1)}$ are dictated by the behavior of $\hu_0^{(1)}$ as $\qv_i \to \vb{0}$. If $\hu_0^{(1)}$ is analytic at the origin, the kernel $\G_0^{(1)}$ is short-ranged and the stationary measure $P_\infty$ stays local at this order; but if it is non-analytic, with $\hu_0^{(1)}\sim q^{\beta}$, the real space kernel decays algebraically, $\G_0^{(1)}\sim |y|^{-(md+\beta)}$, so the stationary measure becomes nonlocal \footnote{We use $q^\alpha$ and $|y|^{\alpha}$ to refer to combinations of the $\qv_i$ and $\yv_i$ such that the exponents add up to $\alpha$; note that non-analyticity is possible also for even $\alpha$ due to the angular dependence in $\qv$.}. We will refer to $\beta$ as the nonlocality exponent.

To present concrete results, summarized in Table~\ref{table} of the End Matter, we now consider a series of well-studied active field theories. In the non-conserved class, we consider Active Model A (AMA)~\cite{caballero-2020}, which sets $\mob = 1$ and ${\lin = 1 - \xi^2\laplacian}$ in \eqref{eq:dynamics}, where $\xi$ is the correlation length. The active nonlinearity is ${\Kc = |\grad\phi|^2}$. Here, both $\Mc_\qv$ and $\Lc_\qv$ are $\order{q^0}$; then $\hu_0^{(1)}$ is analytic, and $\G_0^{(1)}$ is short-ranged, decaying exponentially on a scale set by $\xi$ (except at criticality where this diverges).

In contrast, the conserved class (Model B type) has $\mob = -\laplacian$, ${\lin = 1 - \xi^2\laplacian}$ in \eqref{eq:dynamics}. Crucially, as $\Mc_\qv= q^2$, the denominator of \eqref{eq:u} then vanishes at the origin. As long as the perturbation breaks detailed balance, this leads to a non-analytic $\hu_0^{(1)}$ due to the angular dependence in the $\qv_i$. Inspecting the small $\qv$ scaling in \eqref{eq:u}, we find $\beta = g-2$.
This class includes (in the regime of small activity parameters $\lambda$ and $\zeta$) AMB+, for which $\alpha\Kc = {\lambda\laplacian|\grad\phi|^2} -{\zeta\div(\laplacian\phi\grad\phi)}$, corresponding to $g=4$ and $\nonlin=2$.

In Fig.~\fref[a]{fig:AMAB}, we show $\hu_0^{(1)}$ for AMA and AMB+, contrasting their respective analytic and non-analytic behavior. We may further conclude from the analysis above that $\beta=2$ in AMB+. This holds for any $d>1$; closer inspection reveals that the nonlocality exponent is changed in $d=1$ because the prefactor of the leading-order decay exactly vanishes. We show in~\SM~that in $d=1$, $\hu_0^{(1)}\sim \xi^2q^4$, and thus $\beta=4$ for AMB+ in one dimension.

Next we consider Model AB~\cite{li-2020b}, which combines conservative (B) and non-conservative (A) dynamics; here ${\mob = 1 - \ell^2\laplacian}$ and ${\mob\lin=\aA-\aB\ell^2\laplacian(1 - \xi^2\laplacian)}$, where $\ell$ is the so-called screening length. Nonlinearity is added either to the non-conserved sector ($\Kc_\mathrm{A} = \phi^\nonlin$) or to the conserved one ($\Kc_\mathrm{B}=\laplacian\phi^\nonlin$).
As long as all parameters are finite and non-vanishing, the dynamics as a whole is non-conservative: the analysis outlined above applies, and the stationary measure is local, with $\G_r^{(n)}$ decaying exponentially on a length-scale set by $\ell$ and $\xi$. However, if 
$\aA\to 0$, the linear deterministic dynamics is conservative, and only the noise in \eqref{eq:dynamics} breaks $\phi$-conservation. Since $\Lc_\qv\sim q^2$, this leads to a vanishing denominator at $\qv_i\to \vb{0}$ in \eqref{eq:u}, resulting in a nonlocal stationary measure $P_\infty$, with $\beta=2$. 

The structure of \eqref{eq:u} also appears in the higher-order contributions in \eqref{eq:WKB}, so that the arguments above apply to every other $\Uc_r^{(n)}$. As we show in~\SM, our results are distinct from cases where the stationary measure is nonlocal even in the absence of nonlinearity. This arises in the unscreened limit of  Model AB where $\ell\to \infty$ with $\aA/\ell^2=\mathrm{const.}$~\cite{li-2020b}; in models where the noise conserves the center of mass of the $\phi$ field~\cite{hexner-2017,galliano-2023a,deluca-2024,maire-2025,maire-2025a,maire-2026,gao-2026}; and also in certain long-studied nonequilibrium field theories with anisotropy~\cite{grinstein-1990,garrido-1990,dorfman-1994}.

\begin{figure*}[tbp]
\centering
    \includegraphics[width=\textwidth]{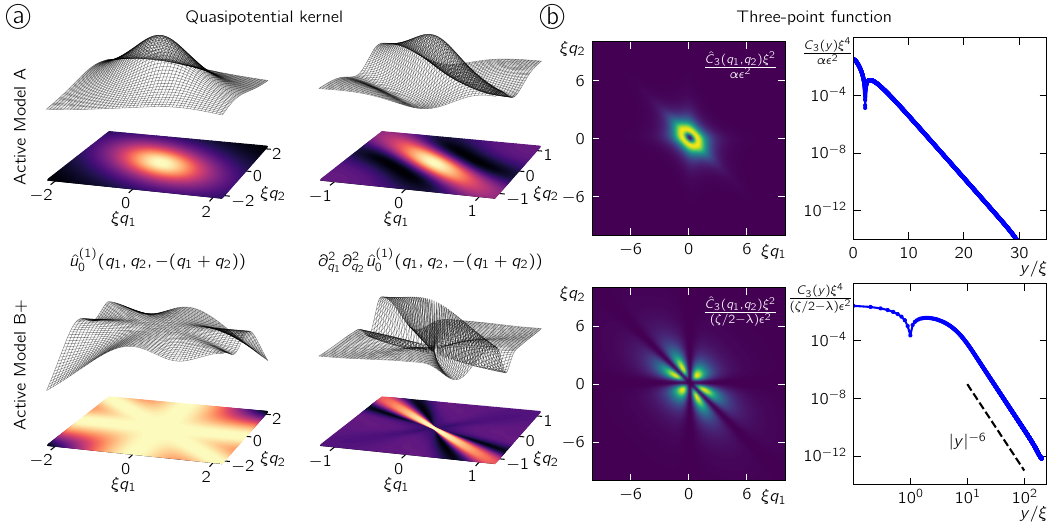}
    \caption{\label{fig:AMAB} Quasipotential kernel and three-point function for the high-temperature phase of AMA (top row) and AMB+ (bottom row) in $d=1$. a) Fourier-space quasipotential kernel $\hu_0^{(1)}$ (left) and its fourth derivative (right). The kernel is analytic at the origin for AMA, while for AMB+, the fourth derivative does not exist at the origin ($\beta=4$). b) Contribution at order $\order{\alpha}$ to the three-point function, related to $\hu_0^{(1)}$ via \eqref{eq:mpt}. In Fourier space (left), we show $\hat C_3$ defined via $\langle\phi_{q_1}\phi_{q_2}\phi_{q_3}\rangle=\hat C_3(q_1,q_2)\dbar_{q_1+q_2+q_3}$, while in real space (right), $C_3(y)=|\langle\phi(0)\phi(y)\phi(y)\rangle|$ is shown. For AMA, $\hat C_3$ is analytic at the origin (top left panel), resulting in exponential decay (with decay length set by $\xi$) of $C_3(y)$ (log-linear plot, top right panel). In AMB+, $\hat C_3$ is non-analytic at the origin (bottom left panel), resulting in algebraic decay in real space (log-log plot, bottom right panel) for length-scales larger than $\xi$, with the expected $C_3\sim |y|^{-(md+\beta)}=|y|^{-6}$ for $\beta=4$, $d=1$, $m=2$.}
\end{figure*}

\sectitle{Breakdown of Landau--Ginzburg theory.}We now show that the nonlocality of $P_\infty$ results in the breakdown of the Landau--Ginzburg approximation. Performing a gradient expansion in \eqref{eq:Uc_G} and assuming isotropy, we formally obtain~\SM:
\begin{equation}\label{eq:GL}
\hspace{-.15em}\Uc_0^{(1)}\!\approx\!\int\!\dd{\xv} \left[\kappa_0\phi^{\nonlin+1}(\xv) + \kappa_2\phi^{\nonlin-1}|\grad\phi|^2 + \cdots\right].
\end{equation}
However, for nonlocal $\Uc[\phi]$ this fails at $\mathcal{O}(\nabla^{\beta'})$ via divergence of the coefficient $\kappa_{\beta'}$. Here $\beta'$ is the smallest even integer greater or equal to the nonlocality exponent $\beta$. In the conserved case, for a nonlinearity with $g$ gradients, we find that $\kappa_{g-2}$ diverges logarithmically in the system size. In the single-field case, the lowest-order active nonlinearity has $g=4$, and the Landau--Ginzburg expansion fails at square-gradient level; this occurs in AMB+ for $d\geq 2$ (while in $d=1$, $\kappa_4$ diverges). The decay of the kernels $\G_r^{(n)}$ is reported in the End Matter.

\sectitle{long-range correlations.}As discussed in~\SM, a general $N$-point correlation function can be found from the stationary measure $P_\infty$, to the desired order in $\alpha$ and $\noise$, using Feynman diagrams. This requires full knowledge of $P_\infty$, including contributions beyond LDT.
Although a generic calculation is cumbersome, we expect that for models with a nonlocal $P_\infty$, any $N$-point correlation function also decays algebraically. We indeed explicitly show in~\SM~that the two-point function for models in the conserved class decays as~\footnote{The AMB+ case in $d=1$ is again special due to an exact cancellation at leading order. We find in this case $|y|^{-8}$~\SM.}:
\begin{align}\label{eq:2pt}
\expval{\phi(\xv)\phi(\xv+\yv)} \sim \alpha^2\noise^m|\yv|^{-[md+2(g-2)]}\,.
\end{align}
Another direct consequence of our analysis is that the leading-order contribution in $\alpha$ to the $(m+1)$-function only depends on $\Uc_0^{(1)}$, and its decay is exactly the same as the one of $\G_0^{(1)}$ (see End Matter). In Fig.~\fref[b]{fig:AMAB}, we show this contribution to the three-point function of AMA and AMB+, demonstrating that in the latter case the nonlocal stationary measure indeed results in algebraic decay of the correlator. The decay of general $\N$-point correlators is reported in the End Matter.

\sectitle{Multiple fields.}Our perturbative procedure to compute the stationary measure is readily generalized to more than one order parameter~\SM. Here, we consider two concrete examples: the Toner--Tu equation (see below) and first the non-reciprocal Cahn--Hilliard (NRCH) model, defined as~\cite{you-2020,saha-2020}:
\begin{subequations}\label{eq:NRCH}
    \begin{align}
    \partial_t\NRCHone &= \laplacian(\lin_{\NRCHone\NRCHone}\NRCHone + \lin_{\NRCHone\NRCHtwo}\NRCHtwo) + \alpha\Kc_\NRCHone +\sqrt{2\noise}\eta_\NRCHone\,,\\
    \partial_t\NRCHtwo &= \laplacian(\lin_{\NRCHtwo\NRCHone}\NRCHone + \lin_{\NRCHtwo\NRCHtwo}\NRCHtwo) +\sqrt{2\noise}\eta_\NRCHtwo\,,
    \end{align}
\end{subequations}
where $\eta_{\NRCHone,\NRCHtwo}$ are independent {\em conserved} noises ({\em i.e.}, ${\mob=-\laplacian}$) and the $\lin_{ij}$ are linear operators. Non-reciprocity means that $\lin_{\NRCHone\NRCHtwo}\neq \lin_{\NRCHtwo\NRCHone}$. The standard NRCH model has
$\Kc_\NRCHone=\laplacian\NRCHone^3$~\cite{you-2020,saha-2020}, but we give results for a general $\Kc_\NRCHone=\order{\nabla^g\NRCHone^{m-s}\NRCHtwo^s}$. Notice that
Eqs.~\eqref{eq:NRCH} break detailed balance even when $\Kc_\NRCHone=0$ but, as the dynamics is then linear, we can still find both its stationary measure and the instanton needed for our perturbative strategy.

Calculating $P_\infty$, 
we find that already in $\Uc_0^{(1)}$, all possible couplings $\NRCHone^{\nonlin+1-l}\NRCHtwo^{l}$ with $0\leq l \leq \nonlin + 1$ are present. Thus, even if the nonlinearity involves $\NRCHone$ only, it even affects the coupling of $\NRCHtwo$ with itself.
Furthermore, the stationary measure is nonlocal, with $\beta=g-2$. Accordingly, 
in the standard NRCH case ($g=2)$, the Landau--Ginzburg expansion \eqref{eq:GL} breaks down at leading order, with $\kappa_0$ logarithmically diverging in the system size.
As in the single-field case, the conservation law is crucial; for a non-conserved analogue of \eqref{eq:NRCH}, the stationary measure remains local at every order~\SM. We finally note that \eqref{eq:2pt} applies to NRCH as well, and so do the generic scalings for kernels and correlators given in the End Matter.

We now turn to a minimal (1D) Toner--Tu model~\cite{toner-1998}:
\begin{subequations}\label{eq:rhop}
 \begin{align}
 &\hspace{-.3em}\partial_t\density =  -\vel\partial_x \pol  + \alpha_\varphi\Kc_\density\,,\\
 &\hspace{-.3em}\partial_t \pol = -(1- \xi^2\partial_x^2) \pol + \prhoc\partial_x\density + \alpha_p\Kc_\pol + \sqrt{2\epsilon}\eta_\pol\,.
 \end{align}
\end{subequations}
Here $\varphi$ encodes density fluctuations, $\pol$ the polar order, and $\eta_\pol$ is non-conservative noise ($\mob=1$). The (advective) nonlinearities are $\Kc_\pol=-\partial_x p^2$ and $\Kc_\density=-\partial_x(\density\pol)$~\cite{toner-1998}.
As we show in~\SM, upon adding generic nonlinearities (see Table~\ref{table}), $P_\infty$ becomes nonlocal in all its kernels, including those involving only $p$. This shows that the conservation of one field ($\varphi$) can induce a nonlocal structure for a coupled but non-conserved one ($p$). Unlike in NRCH, the algebraic decay of the kernels now depends on the fields involved (see Table~\ref{table}). Note that again $\kappa_0$ can diverge, and does so whenever $\alpha_\varphi$ is nonzero.

\sectitle{Discussion.}We have investigated the nonlocality of the stationary measure $P_\infty$ in active field theories, by expanding perturbatively in nonlinearities about the Gaussian limit. Without detailed balance, the stationary measure is generically nonlocal, with perturbatively calculable power laws for kernels and correlators, whenever a conservation law holds at deterministic level. 
This applies in Active Model B+, which describes active phase separation; the Toner--Tu model, which describes flocking; and the non-reciprocal Cahn--Hilliard model, which describes binary systems whose interactions break Newton's third law. In all these models, we showed that $P_\infty$ is nonlocal due to the conservation of one or more fields. Models which lack a conservation law, such as Active Model A, have a local measure at any finite order in our perturbative approach. Nonlocality might still arise in these cases, but if present it lies beyond our perturbation theory. 

The algebraic decay of $\N$-point correlators was already known to be generic in models with conservation laws far from equilibrium ~\cite{grinstein-1990,garrido-1990,grinstein-1991}. Our work identifies such decay as a direct consequence of the nonlocality of $P_\infty$, explaining why, unlike in equilibrium, it emerges deep in the disordered phase, far from criticality.

A crucial finding is that the Landau--Ginzburg expansion of the effective free energy (\emph{e.g.}, \cite{huang-2023}) $\Uc = -\epsilon \ln P_\infty$ fails in conserved active systems: its coefficients diverge at an order that depends on the model at hand. In multiple-field theories such as NRCH and Toner--Tu, a logarithmic blowup can occur at zeroth order in gradients (as it does for the standard nonlinearity of NRCH and the advective nonlinearity of Toner--Tu): there is {\em no bulk Landau expansion} of $\Uc$. In such cases, $\Uc$ is super-extensive, growing (at perturbative level) as $L^d\ln L \gtrsim L^d$. Meanwhile for AMB+ in $d\geq 2$, the Landau--Ginzburg expansion diverges in the square-gradient coefficient, which in equilibrium models (when combined with quartic nonlinearity) controls the surface tension $\sigma$. This divergence presumably foreshadows the breakdown of locality for $P_\infty$ in low-temperature phase coexistence where, as described in Fig.~\ref{fig:lowtemp}, two distinct tensions govern droplet and bubble excitations.

\sectitle{Acknowledgements.}The authors thank Tal Agranov for helpful discussions and a careful reading of the manuscript. FDL acknowledges the support of the University of Cambridge Harding Distinguished Postgraduate Scholars Programme. CN acknowledges the
support of the ANR grant PSAM and of the INP-IRP grant IFAM. This research was supported in part by the UK EPSRC [grant number EP/Z534766/1], and by grant NSF PHY-2309135 to the Kavli Institute for Theoretical Physics (KITP).

\bibliography{../../references}

\appendix
\section*{End Matter}
{
\begin{table*}
\begin{tblr}{width=\textwidth,colspec={
    |X[0.18,c,m]
    |X[0.23,c,m]
    |X[0.35,c,m]
    |X[0.25,c,m]|
}}
 \hline
 Model & Nonlinearity &
 \makecell{Behavior of $\hu_0^{(1)}$ at the origin} &
 \makecell{
    Large-scale decay of $\G_0^{(1)}$
} \\
 \hline \hline
 \makecell{
Active Model A} &
 \makecell{$|\grad\phi|^2$} &
 analytic &
 exponential \\
 \hline
 \makecell{Generic\\conserved case} &
 \makecell{$\order{\nabla^g\phi^m}$} &
 \makecell{non-analytic, $q^{g-2}$} &
 \makecell{$|y|^{-(md+g-2)}$}\\
 \hline
 \makecell{
 Active Model B+} &
 \makecell{$\laplacian|\grad\phi|^2$, $\div(\laplacian\phi\grad\phi)$\\($g=4$, $\nonlin=2$)} &
 \makecell{$d=1$: non-analytic, $q^4$\\$d\geq 2$: non-analytic, $q^2$} &
 \makecell{$d=1:\hspace{1em}|y|^{-6}$\\
 $d\geq 2:\hspace{1em}|y|^{-(2d+2)}$}\\
 \hline
\SetCell[r=2]{m} \makecell{Model AB\\with $\aA\neq 0$}&
\makecell{$\phi^{\nonlin}$} &
 \SetCell[r=2]{m}analytic &
 \SetCell[r=2]{m}exponential\\
 \cline{2-3}
 & \makecell{$\laplacian\phi^{\nonlin}$} &
 &
 \\
 \hline
 \makecell{Model AB\\
 with $\aA=0$} &
 \makecell{$\laplacian\phi^{\nonlin}$} &
 \makecell{non-analytic, $q^2$} &
 \makecell{$|y|^{-(md+2)}$}\\
 \hline
 \SetCell[r=2]{m}NRCH &
 \makecell{$\order{\nabla^g\NRCHone^{\nonlin-s}\NRCHtwo^s}$} &
 non-analytic, $q^{g-2}$ &
 \makecell{$ |y|^{-(md+g-2)}$}\\
 \cline{2-4}
 &
 \makecell{$\laplacian\NRCHone^{3}$\\
 ($g=2$, $\nonlin=3$, $s=0$)} &
 non-analytic, $q^0$ &
 \makecell{$ |y|^{-3d}$}\\
 \hline
  \SetCell[r=2]{m} Toner--Tu
  &
  \makecell{$\Kc_\density=\order{\partial^g_x\density^{\nonlin-s}\pol^{s}}$} &
  \makecell{$\density^{\nonlin+1-l} \pol^{l}$: non-analytic, $q^{s+l+g-2}$} &
  \makecell{$\density^{\nonlin+1-l} \pol^{l}$:
  $|y|^{-(\nonlin+s+l+g-2)}$} \\
  \cline{2-4}
  &
  \makecell{$\Kc_\pol=\order{\partial^g_x \density^{\nonlin-s}\pol^{s}}$} &
  \makecell{$\density^{\nonlin+1-l} \pol^{l}$: non-analytic, $q^{s+l+g-1}$} &
  \makecell{$\density^{s+1-l} \pol^{l}$:
  $|y|^{-(\nonlin+s+l+g-1)}$} \\
\hline
\end{tblr}
\caption{\label{table}Low-$q$ behavior of $\hu_0^{(1)}(\Qv_{m+1})$ and resulting far-field decay of the kernel $\G_0^{(1)}(\Yv_m)$ of \eqref{eq:Uc_G} at linear order in nonlinearity for various active field theories. Theories without conservation laws (AMA and Model AB with $\aA\neq 0$) have analytic $\hu_0^{(1)}$ and local stationary measures, while all those conserving at least one field at deterministic level show non-analytic $\hu_0^{(1)}$ (through the angular dependence in the $\qv_i$), algebraically decaying kernels, and long-range correlations (the $(m+1)$-function decays as $\G_0^{(1)}$). Explicit expressions for $\hu_0^{(1)}$ are provided in~\SM.}
\end{table*}}

\section{Functionals for the computation of the stationary measure}
The functionals of \eqref{eq:Uc_S} are given as follows. For $r=0$,
\begin{align}\label{eq:S0}
&\Sc_0^{(n)}[\phi] ={}\nonumber\\
&{}\quad-\int\ddq\left[\mu^{(n-1)}_{0,-\qv}\Kc_\qv + \sum_{k=1}^{n-1}{\mu^{(n-k)}_{0,-\qv}}\Mc_\qv{\mu^{(k)}_{0,\qv}}\right],
\end{align}
where the generalized chemical potentials associated to the $\Uc_r^{(n)}$ are given by
\begin{align}
\mu^{(n)}_{r,\qv} &= (2\pi)^d\fdv{\Uc_r^{(n)}}{\phi_{-\qv}}\,,
\label{eq:mu}
\end{align}
while for $r>0$, $n>0$ (note that $\Uc_r^{(0)}=0$ for the dynamics of Eq.~\eqref{eq:dynamics}):
\begin{align}\label{eq:Sr}
\Sc_r^{(n)}[\phi] ={}&\int\dd{\qv}
\Biggl[\Mc_\qv\fdv{\mu^{(n)}_{r-1,\qv}}{\phi_{\qv}}+\fdv{\Kc_\qv}{\phi_\qv}\delta_{n,1}\delta_{r,1}\Biggr]\nonumber\\
&{}-\int\ddq\Biggl[\mu^{(n-1)}_{r,-\qv}\Kc_\qv+ 2\sum_{k=1}^{n-1}{\mu^{(n-k)}_{0,-\qv}}\Mc_\qv{\mu^{(k)}_{r,\qv}}\nonumber\\
&{}\quad+\sum_{k=0}^{n}\sum_{s=1}^{r-1}{\mu^{(n-k)}_{r-s,-\qv}}\Mc_\qv{\mu^{(k)}_{s,\qv}}\Biggr]\,.
\end{align}

\subsection{Equilibrium nonlinearities return the correct local free energy}
Let us consider the first-order contribution $\Uc_0^{(1)}$ to the quasipotential for an equilibrium nonlinearity stemming from a local free energy $\tilde \Fc$. For such a $\Kc = -\mob\delta{\tilde\Fc}/\delta{\phi}$, we have in Fourier space
\begin{equation}\label{eq:equilibriumK}
\Kc_\qv = -(2\pi)^d\Mc_\qv\fdv{\tilde \Fc}{\phi_{-\qv}},
\end{equation}
for $\tilde \Fc$ of the form
\begin{align}
\tilde \Fc[\phi] ={}& \int\pddqphi{\nonlin+1} f\brq{\nonlin+1}\,\dbar_{\Qvs{\nonlin+1}}\,,\label{eq:f_Fourier}
\end{align}
where $f\brq{m+1}$ can be chosen to be symmetrised in the $\qv_i$. We thus find from \eqref{eq:equilibriumK}:
\begin{equation}\label{eq:Keq}
\hat K\brq{m} = -(m+1)\Mc_{\Qvs{\nonlin}} f(\Qv_{\nonlin};-\Qvs{\nonlin})\,.
\end{equation}
From \eqref{eq:u}, using the delta function in \eqref{eq:Uc}, this results in
\begin{align}
\hu_0^{(1)}\brq{m+1} &= (m+1)\frac{[f\brq{m+1}\Mc_{\qv_{m+1}}\Lc_{\qv_{m+1}}]^\symm}{\sum_{i=1}^{m+1}\Mc_{\qv_i}\Lc_{\qv_i}}\nonumber\\
&= f\brq{m+1}\,,
\end{align}
which is local by assumption (thus, $\Uc_0^{(1)}=\tilde\Fc$). We then have that $\Kc_\qv = -\Mc_\qv\mu^{(1)}_{0,\qv}$, which from \eqref{eq:S0} and \eqref{eq:Sr} implies that $\Uc_0^{(2)} = 0 = \Uc_1^{(1)}$, and thus $\Uc_0^{(n)} = 0$ for all $n>1$ and $\Uc_r^{(n)}=0$ for all $r>0$. As expected, we thus find $\Uc=\Uc_\mathrm{G}+\alpha\tilde\Fc$ exactly.

\section{Form of \texorpdfstring{$(m+1)$}{(m+1)}-point correlator}
In~\SM, we show that the leading-order ($n=1$) contribution to the connected $(\nonlin+1)$-point function only depends on $\Uc_0^{(1)}$, and is given by:
\begin{align}\label{eq:mpt}
	\Bigl\langle {\textstyle\prod_{i=1}^{\nonlin+1}} \phi_{\qv_i} \Bigr\rangle^\connected &=-\frac{\alpha\noise^{\nonlin} (\nonlin+1)!}{\prod_{i=1}^{\nonlin+1} \Lc_{\qv_i}}\hu_0^{(1)}\brq{\nonlin+1}\,\dbar_{\sum_{i=1}^{\nonlin+1} \qv_i}\nonumber\\
    &\phantom{{}={}}+ O(\alpha^2\epsilon^{(3\nonlin-1)/2})\,.
\end{align}
Thus, for $\Lc_\qv\sim\order{q^0}$, the real-space decay of the $(\nonlin+1)$-point function is the same as the decay of $\G_0^{(1)}$. Similar relations hold in the multiple-field case, so that the decay of $\G_0^{(1)}$ given for those theories in Table~\ref{table} is also the decay of the corresponding $(\nonlin+1)$-point correlator.

\section{Scaling of kernels and correlators for generic \texorpdfstring{$\N$}{N}}
In~\SM, we show for the conserved single-field case and for NRCH that the generic contributions to the stationary measure $\Uc_r^{(n)}$ contain $\mrn=(\nonlin-1)n + 2(1-r)$ fields, with the non-analytic part of their kernels scaling as:
\begin{equation}\label{eq:urn_scale}
\hu_r^{(n)} \sim q^{n(g-2)+dr}\,,
\end{equation}
so that $\G_r^{(n)} \sim |y|^{-[n(g-2)+dr+d(\mrn-1)]}$.

Regarding the correlators, we note the following. Firstly, we find that the contribution of order $\alpha^n$ to the connected $N$-point function scales as $\noise^{[N+n(\nonlin-1)]/2}$. Secondly, for a given order $n$, all the $\Uc_r^{(k)}$ with $k\leq n$ and ${r \leq 1 + (\nonlin-1)n/2 - N/2}$ contribute to the connected $N$-point function.
Finally, the leading-order non-analytic contribution to a generic connected $\N$-point function with $\N>2$ scales as
\begin{equation}
\sim\alpha^{n_*}\epsilon^{[\N+{n_*}(m-1)]/2}q^{{n_*}(g-2)+dr_*}\,,
\end{equation}
with $n_* = \lceil(\N - 2)/(\nonlin - 1)\rceil$ and $r_* = [(m-1){n_*}- (\N-2)]/2$ (note that for odd $\nonlin$, the $\phi\to -\phi$ symmetry of the theory is preserved and odd $\N$-point functions vanish exactly). This translates into a real-space scaling of $\alpha^n\epsilon^{[\N+n_*(m-1)]/2}|y|^{-[n_*(g-2)+dr_*+d(\N-1)]}$. The exception is the two-point function, which scales as \eqref{eq:2pt}.

\end{document}


\title{Supplemental Material: Generic nonlocal statistics of the stationary measure in conserved active systems}

\author{Filippo De Luca
}
\affiliation{DAMTP, Centre for Mathematical Sciences, University of Cambridge, Wilberforce Road, Cambridge CB3 0WA, United Kingdom}

\author{Michael E. Cates
}
\affiliation{DAMTP, Centre for Mathematical Sciences, University of Cambridge, Wilberforce Road, Cambridge CB3 0WA, United Kingdom}

\author{Cesare Nardini
}
\affiliation{Service de Physique de l'\'Etat Condens\'e, CEA, CNRS Universit\'e Paris-Saclay, CEA-Saclay, 91191 Gif-sur-Yvette, France}
\affiliation{Sorbonne Universit\'e, CNRS, Laboratoire de Physique Th\'eorique de la Mati\`ere Condens\'ee, 75005 Paris, France}

\maketitle

\renewcommand{\theequation}{S\arabic{equation}}

\section{Nonlocality of the stationary measure at Gaussian level}
\label{sec:gaussian-nonlocal}
Here, we discuss cases where the stationary measure is nonlocal already for the linear dynamics. We limit ourselves to the single-field case of Eq.~\eqref{meq:dynamics}, setting $\alpha = 0$. The Gaussian stationary measure in Fourier space is given by $P_\infty \propto \exp(-\Uc_\mathrm{G}/\noise)$ with 
\begin{equation}
\Uc_\mathrm{G} = \frac{1}{2}\int \ddq  \Lc_\qv|\phi_\qv|^2\,.
\end{equation}
This stationary measure is nonlocal if $\Lc_\qv$ is non-analytic at the origin. Recall that from \eqref{meq:dynamics}, $\Lc_\qv$ is defined as the ratio between the deterministic operator $\Mc_\qv\Lc_\qv$ and the variance of the noise, $\Mc_\qv$. There are thus two distinct ways of making $\Lc_\qv$ non-analytic:
\begin{enumerate}
\item $\mob$ has more gradients than $\mob\lin$. This occurs whenever the noise respects additional conservation laws: for example, this takes place in conserved dynamics ($\Mc_\qv\Lc_\qv\sim q^2$) with the noise additionally conserving the center of mass ($\Mc_\qv\sim q^4$), see \cite{hexner-2017,galliano-2023a,deluca-2024,maire-2025,maire-2025a,maire-2026,gao-2025}. Another example is non-conservative deterministic dynamics ($\Mc_\qv\Lc_\qv\sim q^0$) with conservative noise ($\Mc_\qv\sim q^2$) (this includes the so-called `unscreened limit' of Model AB, see Ref.~\cite{li-2020b} and Sec.~\ref{sec:MAB}). Both these examples lead to $\Lc_\qv \sim q^{-2}$, and thus a nonlocal real-space kernel $\G_0^{(0)} \sim |y|^{-(d-2)}$.
\item Another way to obtain a nonlocal stationary measure is by allowing the operators $\mob\lin$, $\lin$ to be anisotropic: this allows for angular dependence in the ratio $(\Mc_\qv\Lc_\qv)/\Mc_\qv$. This case is discussed in \cite{grinstein-1990,garrido-1990,dorfman-1994}.
\end{enumerate}

\section{Derivation of the perturbative corrections to the stationary measure}
\label{sec:perturbative}
In this Section, we show how to obtain the corrections $\Uc_r^{(n)}$ to the stationary measure appearing in Eq.~\eqref{meq:WKB}, generalizing the results obtained for the quasipotential ($r=0$) in Ref.~\cite{bouchet-2016}. We refer the reader to the original work for a proof of the existence and uniqueness of the solution for that case.

\subsection{Finite-dimensional dynamics}
We start by deriving our results for the finite-dimensional case and will then proceed to treating field theories in the following Section. Consider an Itô stochastic process of the form
\begin{equation}
\dot\xv(t) = \Kv + \sqrt{2\noise}\Lv\,,
\end{equation}
where $\Lv$ is a Gaussian random vector with mean zero and covariance matrix
\begin{equation}
\expval{\Lv(t)\Lv(t')} = \Mv\,\delta(t-t')\,.
\end{equation}
The corresponding Fokker-Planck equation for the transition probability reads
\begin{equation}\label{eq:FPE}
\partial_t P(\xv,t|\xv_0,t_0) = -\div[\Kv P - \noise\div(\Mv P)]\,.
\end{equation}
We assume the existence of a unique stationary measure $P_\infty(\xv) = \lim_{t\to \infty}P(\xv,t|\xv_0,t_0)$. Working in the weak noise limit $\noise\to 0$, we make the WKB ansatz
\begin{equation}\label{eq:WKB_SM}
P_\infty(\xv) = \frac{1}{Z}\exp(-\sum_{r=0}^\infty\noise^{r-1} \Uc_r(\xv)),
\end{equation}
where by definition $\Uc_0(\xv)$ is the quasipotential, while $Z$ is a normalization factor.

Inserting \eqref{eq:WKB_SM} into \eqref{eq:FPE} at stationarity, and collecting terms of order $\mathcal{O}(\noise^{-1})$, we obtain the Hamilton--Jacobi equation for the quasipotential:
\begin{equation}\label{eq:HJ}
\grad \Uc_0\vdot[\Kv+\Mv\vdot\grad\Uc_0] = 0\,,
\end{equation}
while at order $\mathcal{O}(\noise^{r-1})$, with $r\geq 1$, we find:
\begin{equation}\label{eq:beyondHJ}
    \begin{aligned}
    \grad \Uc_r\vdot[\Kv+2\Mv\vdot\grad\Uc_0] ={}& \div\Kv\delta_{r,1}-\div\div\Mv\delta_{r,2}\\
    &{}+\Mv\boldsymbol{:}\grad\grad \Uc_{r-1}\\
    &{}+2\grad \Uc_{r-1}\vdot\div\Mv\\
    &{}- \sum_{s=1}^{r-1}\grad\Uc_{r-s}\vdot\Mv\vdot\grad\Uc_s\,.
    \end{aligned}
\end{equation}

We now expand the $\Uc_r(\xv)$, as well as the deterministic force $\Kv$ and noise covariance $\Mv$ in powers of the perturbative parameter $\alpha$:
\begin{subequations}\label{eq:alpha_exp}
    \begin{align}
    \Uc_r &= \sum_{n=0}^\infty \alpha^n \Uc_r^{(n)}\,,\\
    \Kv &= \sum_{n=0}^\infty \alpha^n \Kv^{(n)}\,,\\
    \Mv &= \sum_{n=0}^\infty \alpha^n \Mv^{(n)}\,.
    \end{align}
\end{subequations}
Inserting these expansions into \eqref{eq:HJ} and sorting by powers of $\alpha$, we obtain the recursive relations
\begin{subequations}
    \begin{align}
    &\hspace{-.5em}\grad \Uc_0^{(0)}\vdot[\Kv^{(0)}+\Mv^{(0)}\vdot\grad\Uc_0^{(0)}] = 0\,,\\
    &\hspace{-.5em}\grad \Uc_0^{(n)}\vdot\Kv_\mathrm{R}^{(0)}= \Sc_0^{(n)}\left[\Uc_0^{(0)},\ldots,\Uc_0^{(n-1)}\right],\quad n\geq 1\,,\label{eq:Uc_pert}
    \end{align}
\end{subequations}
where
\begin{align}
\Sc_0^{(n)} \coloneqq{}& -\grad \Uc_0^{(0)}\vdot[\Mv^{(n)}\vdot\grad \Uc_0^{(0)} + \Kv^{(n)}]\nonumber\\
&{}-\sum_{k=1}^{n-1}\Big[\grad \Uc_0^{(n-k)}\vdot(\Mv^{(0)}\vdot\grad \Uc_0^{(k)} + \Kv^{(k)})\nonumber\\
&{}\hspace{2.5em} + \sum_{l=0}^{n-k}\grad \Uc_0^{(n-k-l)}\vdot\Mv^{(k)}\vdot\grad \Uc_0^{(l)}\Big]\,,\label{eq:S_pert_finite}
\end{align}
and $\Kv_\mathrm{R}^{(0)} \coloneqq \Kv^{(0)} + 2\Mv^{(0)}\vdot\grad\Uc_0^{(0)}$ is the force appearing in the unperturbed instanton dynamics
\begin{align}
    &\dot\Xv^{(0)} = \Kv_\mathrm{R}^{(0)}\,.
\end{align}
Let us denote $\Xv^{(0)}(t)$ as the solution of this equation with $\Xv^{(0)}(t=-\infty) = \xv_\ast$ and $\Xv^{(0)}(t=0)=\xv$, where $\xv_\ast$ is the fixed point of the dynamics s.t. $\Kv^{(0)}(\xv_\ast) = \vb{0}$. This allows us to solve Eqs.~\eqref{eq:Uc_pert} iteratively by the method of characteristics: if the unperturbed quasipotential $\Uc_0^{(0)}$ is known, the $\Uc_0^{(n)}$ can be obtained order by order via the explicit solution
\begin{equation}\label{eq:Uc_sol}
\Uc_0^{(n)}(\xv) = \int_{-\infty}^0\dd{t}\Sc_0^{(n)}[\Xv^{(0)}(t)]\,.
\end{equation}
This is the result given in Ref.~\cite{bouchet-2016}.

We now derive the analogous expression for $r\geq 1$. Inserting \eqref{eq:alpha_exp} into \eqref{eq:beyondHJ}, we find for all $n$:
\begin{align}\label{eq:Urn}
\grad \Uc_r^{(n)}\vdot\Kv_\mathrm{R}^{(0)}={}{} \Sc_r^{(n)}\left[\{\Uc^{(i)}_j\}^{0\leq i\leq n}_{0\leq j<r};\{\Uc_r^{(i)}\}^{0\leq i< n-1}\right],
\end{align}
with
\begin{align}
\Sc_r^{(n)} ={}& -2\sum_{k=1}^{n}\sum_{l=0}^{n-k}\grad \Uc_r^{(n-k-l)}\vdot \Mv^{(k)}\vdot\grad\Uc_0^{(l)}\nonumber\\
&{}- \sum_{k=1}^{n}\grad \Uc_r^{(n-k)}\vdot\left[\Kv^{(k)}+2\Mv^{(0)}\vdot\grad\Uc_0^{(k)}\right]\nonumber\\
&{}+\div\Kv^{(n)}\delta_{r,1}-\div\div\Mv^{(n)}\delta_{r,2}\nonumber\\
&{}+ \sum_{k=0}^{n}\Mv^{(k)}\boldsymbol{:}\grad\grad \Uc_{r-1}^{(n-k)}\nonumber\\
&{}+2\sum_{k=0}^{n}\grad \Uc_{r-1}^{(n-k)}\vdot\div\Mv^{(k)}\nonumber\\
&{}- \sum_{s=1}^{r-1}\sum_{k=0}^{n}\sum_{l=0}^{n-k}\grad\Uc^{(n-k-l)}_{r-s}\vdot\Mv^{(k)}\vdot\grad\Uc^{(l)}_s\,.\label{eq:Sr_pert_finite}
\end{align}
Analogously to Eq.~\eqref{eq:Uc_pert}, \eqref{eq:Urn} is solved by
\begin{equation}\label{eq:Vc_sol}
\Uc_r^{(n)}(\xv) = \int_{-\infty}^0\dd{t}\Sc_r^{(n)}[\Xv^{(0)}(t)]\,.
\end{equation}
Thus, if $\Uc_0^{(0)}$ and $\Xv^{(0)}(t)$ are known, Eqs.~\eqref{eq:Uc_sol} and \eqref{eq:Vc_sol} allow to iteratively reconstruct all the $\Uc_r^{(n)}$, and thus the measure \eqref{eq:WKB_SM}. Indeed, as is evident from \eqref{eq:Urn}, $\Uc_r^{(n)}$ depends only on previous orders.

\subsection{Field theory (single field)}
The finite-dimensional formalism detailed above is directly extended to field theories like Eq.~\eqref{meq:dynamics}. Note that there we assumed that the noise is additive and unperturbed, while the perturbative force is linear in $\alpha$, so that $\Mv^{(n>0)} = \vb{0} = \Kv^{(n>1)}$. Equations \eqref{eq:Uc_sol} and \eqref{eq:Vc_sol} then become Eq.~\eqref{meq:Uc_S} in the main text.
Promoting gradients to functional derivatives, the functionals $\Sc_r^{(n)}$ appearing there are obtained from Eqs.~\eqref{eq:S_pert_finite} and \eqref{eq:Sr_pert_finite}, yielding
\begin{align}\label{eq:S0_SM}
&\Sc_0^{(n)}[\phi] ={}\nonumber\\
&{}\quad-\int\ddq\left[\mu^{(n-1)}_{0,-\qv}\Kc_\qv + \sum_{k=1}^{n-1}{\mu^{(n-k)}_{0,-\qv}}\Mc_\qv{\mu^{(k)}_{0,\qv}}\right],
\end{align}
and for $r\geq 1$,
\begin{align}\label{eq:Sr_SM}
\Sc_r^{(n)}[\phi] ={}&\int\dd{\qv}
\Biggl[\Mc_\qv\fdv{\mu^{(n)}_{r-1,\qv}}{\phi_{\qv}}+\fdv{\Kc_\qv}{\phi_\qv}\delta_{n,1}\delta_{r,1}\Biggr]\nonumber\\
&{}-\int\ddq\Biggl[\mu^{(n-1)}_{r,-\qv}\Kc_\qv+ 2\sum_{k=1}^{n-1}{\mu^{(n-k)}_{0,-\qv}}\Mc_\qv{\mu^{(k)}_{r,\qv}}\nonumber\\
&{}\quad+\sum_{k=0}^{n}\sum_{s=1}^{r-1}{\mu^{(n-k)}_{r-s,-\qv}}\Mc_\qv{\mu^{(k)}_{s,\qv}}\Biggr]\,,
\end{align}
where the generalized chemical potential is given by
\begin{align}
\mu^{(n)}_{r,\qv} &= (2\pi)^d\fdv{\Uc_r^{(n)}}{\phi_{-\qv}}\,.
\label{eq:mu_SM}
\end{align}
Eqs.~\eqref{eq:S0_SM} and \eqref{eq:Sr_SM} are \eqref{meq:S0} and~\eqref{meq:Sr} in the End Matter and are reported here for convenience.

For the dynamics of \eqref{meq:dynamics}, both the quasipotential $\Uc_0^{(0)} = \Uc_\mathrm{G}$ and the instanton $\Phi^{(0)}(t)$ of the unperturbed dynamics are known. The latter solves:
\begin{align}\label{eq:Phi}
    &\partial_t\Phi^{(0)}_\qv(t) = \Mc_\qv\Lc_\qv\phi_\qv\,,\quad\Phi^{(0)}_\qv(t=0)=\phi_\qv\,,
\end{align}
with $\Phi^{(0)}_\qv(t=-\infty) = 0$ for non-conserved dynamics and $\Phi^{(0)}_\qv(t=-\infty) = \phi_\qv\delta_{\qv,\vb{0}}$ for conserved dynamics. These equations are solved by
\begin{equation}
 \Phi^{(0)}_\qv(t) = e^{\Mc_\qv\Lc_\qv t}\phi_\qv\,.\label{eq:instanton}
\end{equation}

The relation between the number of fields $\mrn$ in $\Sc_r^{(n)}$ (and hence $\Uc_r^{(n)}$) and the order $m$ of the nonlinearity in $\Kc$ reads:
\begin{equation}
\mrn = (m - 1)n+2(1-r).\label{eq:mrn}
\end{equation}
Importantly, for $r > \lceil (m-1)n/2 \rceil$, $\mrn\leq 0$, so that we must then have $\Uc_r^{(n)}\equiv 0$.

We prove \eqref{eq:mrn} by induction. Let us start with $r=0$. For $n=1$ only the first term in Eq.~\eqref{eq:S0_SM} survives, so that $\m(0,1)=m+1$ as required. Now, assume \eqref{eq:mrn} holds for all $\m(0,j)$ with $j\leq n$ and let us show that it holds for $\m(0,n+1)$. In the first term in~\eqref{eq:S0_SM}, there are $m + \m(0,n) - 1 = (m-1)(n+1)+2$ fields, and for each summand in the second term we have $[(m-1)(n+1-k)+2 -1 ]+ [(m-1)k+2 - 1] = (m-1)(n+1)+2$ fields. We conclude that \eqref{eq:mrn} holds for $r=0$. We now turn to $r>0$. For $n=1$, $r=1$, the terms in the first line of \eqref{eq:Sr_SM} both have $m-1$ fields ($\m(0,1)-2=m-1$), and the only surviving term in the second line also has $m+\m(0,0)-1=m-1$ fields, so that as required $\m(1,1)=m-1$. For the induction step, assume that Eq.~\eqref{eq:mrn} holds for all the previous iteration steps, and notice that going from $n$ to $n+1$ adds $m$ fields in each of the terms of \eqref{eq:Sr_SM}; while $r\to r+1$ results in 2 fewer fields everywhere; thus, \eqref{eq:mrn} holds for all $r,n$.

Due to translational invariance, we can always write:
\begin{align}
\Sc_r^{(n)} &= \int\pddqphi{\mrn}\hat s_r^{(n)}\brq{\mrn}\,\dbar_{\Qvs{\mrn}}\,,\label{eq:s_Fourier}\\
\Uc_r^{(n)} &= \int\pddqphi{\mrn}\hu_r^{(n)}\brq{\mrn}\,\dbar_{\Qvs{\mrn}}\,.\label{eq:u_Fourier}
\end{align}
The functions $\hat s_r^{(n)}\brq{\mrn}$ and $\hu_r^{(n)}\brq{\mrn}$ can be assumed to be fully symmetrized in the momenta $\qv_i$. In our setting, because $\Phi^{(0)}(t)$ is known explicitly from \eqref{eq:instanton}, the time integral in Eq.~\eqref{meq:Uc_S} connecting \eqref{eq:s_Fourier} and \eqref{eq:u_Fourier} is particularly easy to perform. We find:
\begin{equation}\label{eq:s_to_u}
\hu_r^{(n)}\brq{\mrn} = \frac{\hat s_r^{(n)}\brq{\mrn}}{\sum_{i=1}^{\mrn}\Mc_{\qv_i}\Lc_{\qv_i}}\,.
\end{equation}
The functions $\hat s_r^{(n)}\brq{\mrn}$ are obtained as follows. First, the generalized potential of \eqref{eq:mu_SM} is given by
\begin{align}\label{eq:chempot_n}
\mu^{(n)}_{r,-\qv} = \mrn\int&\pddqphi{\mrn-1}{}\hu_r^{(n)}(\Qv_{\mrn-1};\qv)\cdot{}\nonumber\\
&\quad{}\cdot\dbar_{\qv+\Qvs{\mrn-1}}\,.
\end{align}
Using this in \eqref{eq:S0_SM} and~\eqref{eq:Sr_SM}, we find
\begin{widetext}
    \begin{align}
    \hat s_0^{(n)}\brq{\m(0,n)} ={}& -\m(0,n-1)\left[\hat K\brq{m}\hu_0^{(n-1)}(\Qv_{m+1\to\m(0,n)};\Qvs{m})\right]^\symm\nonumber\\
    &{}\hspace{-4em}-\sum_{k=1}^{n-1}\m(0,n-k)\m(0,k)\Biggl[\Mc_{\Qvs{\m(0,k)-1}}\hu_0^{(k)}(\Qv_{\m(0,k)-1};-\Qvs{\m(0,k)-1})\,\hu_0^{(n-k)}(\Qv_{\m(0,k)\to\m(0,n)};\Qvs{\m(0,k)-1})\Biggr]^\symm,\label{eq:sn}
    \end{align}
    where $\Qv_{i\to j}=(\qv_i,\ldots,\qv_{j})$. For $r\geq 1$,
    \begin{align}
    \hat s_r^{(n)}\brq{\m(r,n)} ={}&{} -\m(r,n-1)\left[\hat K\brq{m}\hu_r^{(n-1)}(\Qv_{m+1\to\mrn};\Qvs{m})\right]^\symm\nonumber\\
    &{}\hspace{-3em} -2\sum_{k=1}^{n-1}\m(0,n-k)\m(r,k)\Biggl[\Mc_{\Qvs{\m(r,k)-1}}\hu_r^{(k)}(\Qv_{\m(r,k)-1};-\Qvs{\m(r,k)-1})\,\hu_0^{(n-k)}(\Qv_{\m(r,k)\to\m(r,n)};\Qvs{\m(r,k)-1})\Biggr]^\symm\nonumber\\
    &{}\hspace{-3em} -\sum_{s=1}^{r-1}\sum_{k=0}^{n}\m(r-s,n-k)\m(s,k)\Biggl[\Mc_{\Qvs{\m(s,k)-1}}\hu_s^{(k)}(\Qv_{\m(s,k)-1};-\Qvs{\m(s,k)-1})\,\hu_{r-s}^{(n-k)}(\Qv_{\m(s,k)\to\m(r,n)};\Qvs{\m(s,k)-1})\Biggr]^\symm\nonumber\\
    &{}\hspace{-3em}+\m(r-1,n)[\m(r-1,n)-1]\,\int\ddq\Mc_\qv\hu_{r-1}^{(n)}(\Qv_{\m(r-1,n)-2};\qv,-\qv)+m\int\ddq\hat K(\Qv_{m-1};\qv)\delta_{r,1}\delta_{n,1}\,.
    \label{eq:srn}
    \end{align}
\end{widetext}
Together with \eqref{eq:s_to_u}, these equations give an iterative procedure to reconstruct the kernels $\hu_r^{(n)}$. For the linear correction to the quasipotential ($r=0$, $n=1$), $\m(0,1)=m+1$, and Eq.~\eqref{eq:sn} becomes simply:
\begin{align}
&\hat s_0^{(1)}\brq{m+1} =-\left[\hat K\brq{m}\Lc_{\qv_{m+1}}\right]^\symm,\label{eq:s1}
\end{align}
which with \eqref{eq:s_to_u} yields Eq.~\eqref{meq:u} of the main text.

\subsection{Real-space stationary measure}
\label{sec:u_real}
Here, we show how to express the $\Uc_r^{(n)}$ in real space. We apply the Fourier transform
\begin{align}\label{eq:FT}
    \phi_\qv = \FT\phi(\xv)&= \int\dd{\xv} e^{-i\qv\vdot\xv}\phi(\xv)
\end{align}
to the fields in \eqref{eq:u_Fourier} and use the delta function to obtain:
\begin{align}
\Uc_r^{(n)}[\phi]
={}&\int\prod_{i=1}^\mrn[\dd{\yv_i}\phi(\yv_i)]\pddq{{\mrn-1}}\cdot{}\nonumber\\
&{}\cdot{} e^{-i\sum_{i=1}^{\mrn-1}\qv_i\vdot(\yv_i-\yv_{\mrn})}\nonumber\\
&{}\cdot{} \hu_r^{(n)}\!\left(\Qv_{\mrn-1};-\Qvs{\mrn-1}\right)\!,\nonumber\\
={}&\int\!\dd{\xv}\phi(\xv)\hspace{-1em}\prod_{i=1}^{\mrn-1}\hspace{-1em}\left[\dd{\yv_i}\phi(\xv+\yv_i)\right] \G_r^{(n)}(\Yv_{\mrn-1})\,,\label{eq:U_to_G}
\end{align}
where in the last line we have renamed $\yv_\mrn\to \xv$, shifted $\yv_i \to \yv_i-\xv$ for $i=1,\ldots, \mrn-1$, and introduced the real-space kernel:
\begin{equation}
\G_r^{(n)}(\Yv_{\mrn-1}) = \FT^{-1}\hu_r^{(n)}\!\left(-\Qv_{\mrn-1};\Qvs{\mrn-1}\right).
\end{equation}
Thus, if $\hu_r^{(n)}\!\left(-\Qv_{\mrn-1};\Qvs{\mrn-1}\right)$ is non-analytic at the origin, $\G_r^{(n)}(\Yv_{\mrn-1})$ is long-ranged. For $\hu_r^{(n)}$ non-analytic with $\hu_r^{(n)}\sim q^\beta$, so that the $\beta$-th derivative does not exist, we have $\G_r^{(n)} \sim |y|^{-[(\mrn-1)d+\beta]}$.

\subsection{Explicit expressions}
\subsubsection{Active Model A}
We now derive explicit expressions for $\hu_0^{(1)}$ for the single-field theories of the main text. We start from AMA, where $\Mc_\qv=1$, $\Lc_\qv=1+\xi^2q^2$, and $\hat K(\Qv_2) = -\qv_1\vdot\qv_2$. Equation \eqref{meq:u} gives
\begin{equation}\label{eq:uAMA}
\hu_{0}^{(1)}(\Qv_3) = \frac{[(\qv_1\vdot\qv_2)(1+\xi^2q_3^2)]^\symm}{\sum_{i=1}^3(1+\xi^2q_i^2)}.
\end{equation}
This is clearly analytic at the origin, resulting in a local $\Uc_0^{(1)}$.

\subsubsection{Active Model B+}
In AMB+, $\Mc_\qv=q^2$, $\Lc_\qv=1+\xi^2q^2$, and $\alpha\hat K(\Qv_2) = \lambda(\qv_1+\qv_2)^2\qv_1\vdot\qv_2 - \zeta (\qv_1+\qv_2)\vdot[q_1^2\qv_2]^\symm$. With $\qv_3=-\Qvs{2}$, \eqref{meq:u} gives
\begin{equation}\label{eq:uAMB}
\alpha\hu_{0}^{(1)}(\Qv_3) = \frac{[(\qv_1\vdot\qv_2)q_3^2(\lambda(1+\xi^2q_3^2)+\zeta(1+\xi^2q_1^2))]^\symm}{\sum_{i=1}^3q_i^2(1+\xi^2q_i^2)}.
\end{equation}
This is non-analytic at the origin, leading to a nonlocal $\Uc_0^{(1)}$. We have that $\hu_0^{(1)}\sim q^2$ ($\beta=2$), so that the non-analyticity manifests itself in non-existing second and higher derivatives at the origin.

The case $d=1$ is special. To see why, note that the numerator in \eqref{eq:uAMB} reads:
\begin{equation}
\alpha \hat s_{0}^{(1)}(\Qv_2;-\qv_1-\qv_2) = \frac{2}{3}(\lambda+\zeta)[(\qv_1\vdot\qv_2)^2-q_1^2q_2^2] + \order{\xi^2q^6},
\end{equation}
so that the leading-order term vanishes only in $d=1$. In this case, we obtain from \eqref{eq:uAMB}:
\begin{equation}
    \begin{aligned}
    &\alpha\hu_{0}^{(1)}(q_1,q_2,-q_1-q_2) =\\
    &\qquad\frac{\xi^2(\lambda-\zeta/2)q_1^2q_2^2(q_1^2+q_2^2+2q_1q_2)}{2(q_1^2+q_2^2+q_1q_2)+2\xi^2(q_1^2+q_2^2+q_1q_2)^2}\,.
    \end{aligned}
\end{equation}
Thus, we find that due to the cancellation in $d=1$, $\hu_0^{(1)}\sim \xi^2q^4$, so that fourth derivatives at the origin do not exist, as shown in Fig.~\fref[a]{mfig:AMAB}. (Further note that for $\lambda=\zeta/2$, the AMB+ nonlinearity vanishes identically in $d=1$, and accordingly, so does $\hu_0^{(1)}$).

\subsubsection{General conservative models}
More generally, for an active nonlinearity with $m$ fields and $g$ gradients, for the conserved case $\Mc_\qv=q^2$ and $\Lc_\qv=1+\xi^2q^2$ we find from \eqref{meq:u} that $\hu_0^{(1)}\sim q^{g-2}$. By induction, we can show that $\hu_0^{(n)}\sim q^{n(g-2)}$: indeed, if this is true up to order $n-1$, from \eqref{eq:sn} we see that $\hat s_0^{(n)}$ scales as $q^{(g-2)n+2}$; with \eqref{eq:s_to_u} we obtain the sought result.

Again by induction, we can then show for all $r,n$ that the non-analytic part of the kernel scales as:
\begin{equation}\label{eq:urn_scale_SM}
\hu_r^{(n)} \sim {q^{n(g-2)+dr}}.
\end{equation}
For $\hu_1^{(1)}$, this is true because in the last line of \eqref{eq:srn} both terms go like $q^{g+d}$, so that with \eqref{eq:s_to_u}, $\hu_1^{(1)}\sim {q^{g-2+d}}$. Applying the induction step to \eqref{eq:srn} confirms the sought result.

Note that for a coupling with fixed number of fields $\N$, there are contributions from $r,n$ with $r=1+(m-1)n/2-\N/2$,
s.t. $\hu_r^{(n)} \sim q^{n(g-2+d(m-1)/2)-d(\N-2)/2}$. This implies that the contributions from higher orders $r, n$ to a given $\N$-body interaction decay faster and faster for conservative nonlinearities ($g\geq 2$).

\subsubsection{Model AB}
\label{sec:MAB}
In Model AB, $\Mc_\qv = 1 + \ell^2 q^2$, $\Lc_\qv = [\aA + \aB(1+\xi^2q^2) q^2\ell^2]/(1+q^2\ell^2)$, and $\hat K_\mathrm{A}(\Qv_m) = 1$, while $\hat K_\mathrm{B}(\Qv_m) = -(\Qvs{m})^2$. We find from \eqref{meq:u}:
\begin{equation}\label{eq:u_MAB}
\hu_0^{(1)}\brq{\nonlin+1} =-\frac{\left[\hat{K}_{\mathrm{A,B}}\brq{\nonlin}\,\frac{\aA + \aB(1+\xi^2q_{m+1}^2) q_{m+1}^2\ell^2}{1+q_{m+1}^2\ell^2}\right]^\symm}{\sum_{i=1}^{\nonlin + 1} [\aA + \aB(1+\xi^2q_i^2) q_i^2\ell^2]}\,,
\end{equation}
so that for $\aA, \ell$ finite and non-zero, $\Uc_0^{(1)}$ is local. There are two limits where the denominator can vanish: the first is the limit $\aA\to 0$, where $\Lc_\qv \sim q^2$ and \eqref{eq:u_MAB} becomes
\begin{equation}
\hu_0^{(1)}\brq{\nonlin+1} =-\frac{\left[\hat{K}_{\mathrm{A,B}}\brq{\nonlin}\,\frac{(1+\xi^2q_{m+1}^2) q_{m+1}^2\ell^2}{1+q_{m+1}^2\ell^2}\right]^\symm}{\sum_{i=1}^{\nonlin + 1} [(1+\xi^2q_i^2) q_i^2\ell^2]}\,,
\end{equation}
This is non-analytic, resulting in a nonlocal $\Uc_0^{(1)}$ with $\beta=g$. This limit corresponds to the unperturbed dynamics being conservative at the deterministic level.

The second special limit is the `unscreened limit' $\ell \to \infty$ with $\aA/\ell^2$ finite, where $\Lc_{\qv}\sim q^{-2}$. In this limit, the noise is conservative, while the deterministic dynamics isn't. Then, as discussed in Sec.~\ref{sec:gaussian-nonlocal}, already the Gaussian stationary measure is nonlocal.

\subsection{UV behavior}
For large $\qv$, the expression for $\hu_0^{(1)}$ in \eqref{meq:u} diverges as $q^{g-g_\mob}$, where $g$ is the number of gradients in $\Kc$ and $g_\mob$ the number of gradients in $\mob$. Accordingly, its Fourier transform $\G_0^{(1)}$ is UV-divergent. This is no surprise: even for an equilibrium $\phi^4$ nonlinearity ($\Kc=\alpha\mob\phi^3$), which corrects the free energy by
\begin{equation}\label{eq:phi4}
\Uc_0^{(1)} = \frac{1}{4}\int\dd{\xv}\phi(\xv)^4\,,
\end{equation}
we read off from \eqref{meq:Uc} and \eqref{meq:Uc_G} that $\hu_0^{(1)}\brq{4} = 1/4$ and $\G_0^{(1)}(\Yv_3) = \delta(\yv_1)\delta(\yv_2)\delta(\yv_3)/4$. Thus, the UV divergence of $\G_0^{(1)}$ is due to its singular structure at small distances, reflecting the locality of functionals like \eqref{eq:phi4}. Throughout this work, we are instead interested in the large-scale, cutoff-independent behavior of the $\G_r^{(n)}$, which is entirely dictated by the behavior of $\hu_r^{(n)}$ close to the origin~\cite{lighthill-1958}.

\subsection{Local approximation of nonlocal stationary measure}
We now show how to formally obtain a local approximation of nonlocal measures and demonstrate at which order this approximation fails. Let us suppose we have $P_\infty \propto \exp(-\Uc/\noise)$ with:
 \begin{align}\label{eq:Uc_G2}
    &\Uc\!=\!\int\!\dd{\xv} \phi(\xv) \prod_{i=1}^{\nonlin} [\dd{\yv_i}\phi(\xv+\yv_i)] \G(\Yv_{\nonlin})\,,
\end{align}
with kernel $\G(\Yv_{\nonlin})$. Taylor expanding $\phi(\xv+\yv_i)=\phi(\xv)+\yv_i\vdot\grad\phi(\xv)+\yv_i\yv_i\boldsymbol{:}\grad\grad\phi(\xv)/2$ and integrating by parts gives:
\begin{widetext}
    \begin{equation}
        \begin{aligned}
        \Uc ={}& \int\!\dd{\xv}\Biggl[\phi^{\nonlin+1}(\xv)\int\prod_{i=1}^\nonlin \dd{\yv_i}\G(\Yv_\nonlin)+\phi^{\nonlin}(\xv)\sum_{j=1}^{\nonlin}\grad\phi(\xv)\vdot \int\prod_{i=1}^\nonlin \dd{\yv_i}\sum_{j=1}^\nonlin\vb{y}_j\,\G(\Yv_\nonlin)\,\\
        &{}\quad+\phi^{\nonlin-1}\grad\phi\grad\phi\boldsymbol{:}\int\prod_{i=1}^\nonlin \dd{\yv_i}\sum_{j=1}^\nonlin\Biggl(\sum_{k\neq j}\frac{\yv_j\yv_k}{2}-\nonlin\frac{\yv_j\yv_j}{2}\Biggr)\G(\Yv_\nonlin)+ \order{\nabla^4}\Biggr].
        \end{aligned}
    \end{equation}
\end{widetext}
Due to isotropy, the $\yv$ integral in the second term of the first line vanishes, while the integral in the second line must be proportional to the identity matrix. We thus obtain the Landau--Ginzburg form
\begin{align}
    \Uc
    &= \int\!\dd{\xv}\left[\kappa_0\phi^{\nonlin+1}(\xv)+\kappa_2\phi^{\nonlin-1}|\grad\phi|^2\right],
\end{align}
with the moments
\begin{subequations}
    \begin{align}
    \kappa_0 &\coloneqq \int\prod_{i=1}^\nonlin \dd{\yv_i}\G(\Yv_\nonlin)\,,\\
    \kappa_2 &\coloneqq \frac{1}{d}\int\prod_{i=1}^\nonlin \dd{\yv_i}\sum_{j=1}^\nonlin\left(\sum_{k\neq j}\frac{\yv_j\vdot\yv_k}{2}-\nonlin\frac{y_j^2}{2}\right)\G(\Yv_\nonlin)\,.
    \end{align}
\end{subequations}
Now, if the Fourier transform of $\G$ is non-analytic with $\hu \sim q^\beta$, we have that $\G \sim |y|^{-(\nonlin d+\beta)}$. In order to have an infrared-convergent $\kappa_n$, we thus require $n < \beta$. For $n = \beta$, the divergence is logarithmic with the system size; for $n > \beta$, it is algebraic.

From \eqref{eq:Uc_G2}, taking $\phi(\xv)=\bar\phi=\mathrm{const.}$, we see that $\Uc = \bar\phi^{\nonlin+1} L^d \cdot \kappa_0$. Hence if $\kappa_0$ is finite, $\Uc$ is extensive; but if $\kappa_0\sim \log L$, as is the case whenever $\beta = 0$, $\Uc$ is logarithmically superextensive.

\section{Derivation of stationary \texorpdfstring{$\N$}{N}-point functions}
\label{sec:correlators}
\subsection{Diagrammatic derivation from the stationary measure}
\label{sec:diagrams_correlators}
We now consider the relation between the perturbative expansion of the stationary measure and the perturbative calculation of $\N$-point correlation functions. These are defined as:
\begin{equation}
F_\N\brq{N} \coloneqq \prod_{i=1}^\N\phi_{\qv_i}\,.\label{eq:npt}
\end{equation}
Due to translational invariance, we can write the expectation value of the $\N$-point function with respect to the stationary measure as:
\begin{equation}
\expval{F_\N\brq{N}} \eqqcolon \hat C_\N\brq{\N}\dbar_{\Qvs{N}}\,,
\end{equation}
and from \eqref{eq:FT}, we find the real-space $\N$-point function
\begin{equation}
\expval{\phi(\xv)\prod_{i=1}^{\N-1}\phi(\xv+\yv_i)} = C_\N(\Yv_{\N-1})\,,
\end{equation}
with
\begin{equation}
C_\N(\Yv_{\N-1}) = \FT^{-1}\hat C_\N(\Qv_{\N-1};-\Qvs{\N-1})\,.
\end{equation}
Analogously to Sec.~\ref{sec:u_real}, if $\hat C_\N$ is non-analytic at the origin, $C_\N$ is long-ranged.

Let us derive the scaling of the $\N$-point functions with $\alpha$ and $\noise$. To this goal, we rescale the field as $\phi \to \phi/\noise^{1/2}$ in Eq.~\eqref{meq:dynamics}, resulting in:
\begin{equation}\label{eq:dynamics_rescaled}
    \partial_t\phi(\xv,t) = -\mob\lin\,\phi + \alpha\epsilon^{(m-1)/2}\Kc[\phi] + \sqrt{2}\,\eta\,,
\end{equation}
Quantities in this rescaled theory can only depend on the combination $\alpha \noise^{(m-1)/2}$. Once the rescaling is undone, we then find that the $\N$-point functions at order $\alpha^{k}$ can only depend on $\noise^{\N/2+k(m-1)/2}$.

We now show how to obtain the expectation value of the $\N$-point function of \eqref{eq:npt} with respect to the measure of \eqref{eq:WKB_SM}. This explicitly reads:
\begin{align}
\expval{F_\N} &{}= \frac{\int\mathcal{D}\phi\,F_\N e^{-{\Uc} [\phi]/\noise}}{\int\mathcal{D}\phi\,e^{-{\Uc} [\phi]/\noise}}\nonumber\\
&{}= \frac{\int\mathcal{D}\phi\,F_\N e^{-\Uc_\mathrm{G}/\noise}e^{-\sum_{n=1}^\infty\alpha^n \sum_{r=0}^\infty \noise^{r-1}\Uc_r^{(n)}}}{\int\mathcal{D}\phi\,e^{-\Uc_\mathrm{G}/\noise}e^{-\sum_{n=1}^\infty\alpha^n\sum_{r=0}^\infty \noise^{r-1}\Uc_r^{(n)}}}\,,
\end{align}
where $\mathcal{D}\phi$ is the path integral measure. We now rescale $\phi \to \phi/\noise^{1/2}$, resulting in:
\begin{align}
\label{eq:operator_expval}
\expval{F_\N} &{}= \noise^\frac{\N}{2}\frac{\int\mathcal{D}\phi\,F_\N e^{-\Uc_\mathrm{G}}e^{-\sum_{n=1}^\infty \sum_{r=0}^\infty \alpha^n\noise^{r-1+\mrn/2}\Uc_r^{(n)}}}{\int\mathcal{D}\phi\,e^{-\Uc_\mathrm{G}}e^{-\sum_{n=1}^\infty\sum_{r=0}^\infty \alpha^n\noise^{r-1+\mrn/2}\Uc_r^{(n)}}}\nonumber\\
&{}= \noise^\frac{\N}{2}\frac{\int\mathcal{D}\phi\,F_\N e^{-\Uc_\mathrm{G}}e^{-\sum_{n=1}^\infty \sum_{r=0}^\infty [\alpha\noise^{(m-1)/2}]^n\Uc_r^{(n)}}}{\int\mathcal{D}\phi\,e^{-\Uc_\mathrm{G}}e^{-\sum_{n=1}^\infty\sum_{r=0}^\infty [\alpha\noise^{(m-1)/2}]^n\Uc_r^{(n)}}}\,,
\end{align}
where we have used \eqref{eq:mrn} (in fact, the argument of \eqref{eq:dynamics_rescaled} can be used with \eqref{eq:operator_expval} as an alternative derivation of \eqref{eq:mrn}).

We can now safely expand \eqref{eq:operator_expval} for small $\alpha\noise^{(m-1)/2}$. This step follows the standard procedure of perturbing around a Gaussian measure (see for example Chapter~5.2 in \cite{kardar-2007}). Once the rescaling is undone, \eqref{eq:operator_expval} becomes:
\begin{align}
    \expval{F_\N} ={}& \!\sum_{k=0}^\infty \frac{(-1)^k}{k!}\expval{\!\left(\sum_{n=1}^\infty\alpha^n\,\sum_{r=0}^{\mathclap{\lceil(m-1)/n\rceil}}\,\noise^{r-1} \Uc^{(n)}_r\right)^k\!F_\N}^\connected_0\,.\label{eq:observable}
\end{align}
where $\expval{\cdot}_0$ denotes the average with respect to $\Uc_\mathrm{G}$ and $\expval{\cdot}_0^\connected$ only considers connected contributions (\emph{i.e.}, exclusively taking into account the contractions that connect each of the $k$ copies of $\sum_{n=1}^\infty\alpha^n\sum_{r=0}^\infty\noise^{r-1} \Uc^{(n)}_r$ with each other and with $F_\N$).

\subsection{Feynman rules}
\label{sec:feynman}
Equation~\eqref{eq:observable} can be codified into diagrammatic Feynman rules. Although these are constructed in the standard textbook way, the fact that the stationary measure contains infinitely many couplings implies that care should be taken in computing the correlation functions. We thus explicitly recall the rules that we use to construct the Feynman diagrams, for the reader's convenience.

Say we are interested in the connected $\N$-point function in Fourier space, $\expval{F_\N}^\connected$, to order $k$ in $\alpha$. To calculate it, we have to consider the couplings coming from all the $\Uc_r^{(n)}$ with $n\leq k$. Each of those couplings contributes a vertex with $\mrn$ legs.
\begin{enumerate}
    \item To calculate the contribution of order $\alpha^k$, we consider partitions of $k$ into integers as explained below; each integer $n$ in every partition represents a coupling from the $\Uc_r^{(n)}$, where $r\leq \lceil(m-1)n/2\rceil$. For example, a contribution to order $\alpha^3$ has to consider diagrams corresponding to the partitions $(1,1,1)$, $(2,1)$, and $(3)$, thus containing either three $\Uc_r^{(1)}$ vertices, or one $\Uc_r^{(2)}$ vertex and one $\Uc_r^{(1)}$ vertex, or one $\Uc_r^{(3)}$ vertex. Each vertex has $\mrn$ legs.
    \item We consider all connected diagrams with the number and type of vertices from Step 1, with $\N$ external legs. Each line is labeled by a momentum $\qv$; these must be chosen taking into account that at every vertex the total momentum is conserved.
    \item Each line contributes a factor coming from its Gaussian propagator, $\noise/\Lc_\qv$. Each vertex contributes a factor $\alpha^n\noise^{r-1}\hu_r^{(n)}\brq{\mrn}$.
    \item Each internal momentum $\tilde\qv_i$ that survives after considering momentum conservation is integrated over with measure $\ddbar\tilde\qv_i$.
    \item Finally, each diagram is multiplied with a factor $(-1)^s/s!$, where $s$ is the number of vertices in the diagram, with its topological multiplicity, and with $\dbar_{\Qvs{N}}$ ensuring overall momentum conservation.
\end{enumerate}

\subsection{Tree-level \texorpdfstring{$\N$}{N}-point functions}
\label{sec:treelevel}
From the Feynman rules just derived, it is evident that only certain $\N$-point functions can be calculated with the sole knowledge of the quasipotential $\Uc_0$ (\emph{i.e.}, without the beyond-LDT contributions $\Uc_r$). Indeed, these are only the $\m(0,n)$-point functions at order $\alpha^n$ (which are tree-level diagrams). To see why, note that the beyond-LDT contributions from $\Uc_r^{k}$ with $r\geq 1,k\leq n$ that would enter in Step 1 have $\m(r,k)<\m(0,n)$ legs and hence do not contribute to the $\m(0,n)$-point function to this order. Using the Feynman rules above, we explicitly find:
\begin{align}\label{eq:mpt_implicit}
	\left\langle F_{\m(0,n)} \right\rangle^\connected =&{}-\frac{\alpha^n\noise^{\m(0,n)-1} {\m(0,n)}!}{\prod_{i=1}^{\m(0,n)} \Lc_{\qv_i}}\hu_0^{(n)}\brq{\m(0,n)}\,\dbar_{\Qvs{\m(0,n)}}\nonumber\\&{} + O(\alpha^2)\,,
\end{align}
which for $n=1$ results in Eq.~\eqref{meq:mpt}.

In contrast, any $\N$-point function that requires loop diagrams also needs to consider the beyond-LDT contributions, since closing a loop in a quasipotential vertex (which removes two legs) results in a contribution to the same order in $\noise,\alpha$ as a loopless vertex from $\Uc_1^{(n)}$, due to $\m(1,n) = \m(0,n) - 2$.
An explicit example for a loop calculation is shown in the next Section.

\subsection{Explicit calculation of correlations in AMB+}
\label{sec:loop}
We now give explicit results for the lowest-order corrections to the correlators for AMB+, using the rules of Sec.~\ref{sec:feynman}. We will need the vertices to order $\alpha^2$. Since we have $m=2$, these are:
\begin{equation}\label{eq:AMB_vertices}
\begin{tikzpicture}[baseline={(0,-0.1)}]
    \begin{feynman}
        \vertex (a) at (0,0);
        \vertex[dot, minimum size=3.5pt, label={[label distance=.2em]right:$\hat{u}_0^{(1)}$}] (b) at (0.5,0) {};
        \vertex (c) at (0.933,0.5);
        \vertex (d) at (0.933,-0.5);
        \diagram* {
            (a) -- (b),
            (b) -- (c),
            (b) -- (d)
        };
    \end{feynman}
\end{tikzpicture},\quad
\begin{tikzpicture}[baseline={(0,-0.1)}]
    \begin{feynman}
        \vertex[red, dot, minimum size=3.5pt, label={[label distance=.1em]below:$\hat{u}_1^{(1)}$}] (a) at (0,0) {};
        \diagram* {
            (a)
        };
    \end{feynman}
\end{tikzpicture},\quad
\begin{tikzpicture}[baseline={(0,-0.1)}]
    \begin{feynman}
        \vertex (a) at (0,0);
        \vertex[empty dot, minimum size=3.5pt, label={[label distance=.1em]below right:$\hat{u}_0^{(2)}$}] (b) at (0.5,0) {};
        \vertex (c) at (1.0,0);
        \vertex (d) at (0.5,0.5);
        \vertex (e) at (0.5,-0.5);
        \diagram* {
            (a) -- (b),
            (b) -- (c),
            (b) -- (d),
            (b) -- (e)
        };
    \end{feynman}
\end{tikzpicture},\quad
\begin{tikzpicture}[baseline={(0,-0.1)}]
    \begin{feynman}
        \vertex[red, empty dot, minimum size=3.5pt, label=below:$\hat{u}_{1}^{(2)}$] (m) at (0, 0) {};
        \vertex (a) at (-0.5,0);
        \vertex (b) at ( 0.5,0);
        \diagram* {
            (a) -- (m),
            (m) -- (b),
        };
    \end{feynman}
\end{tikzpicture}\,.
\end{equation}
Note that the one-field vertex $\hu_1^{(1)}$ vanishes identically in AMB+ due to the conservation law.

As explained in Sec.~\ref{sec:treelevel}, the $(m+1)$-function only has a tree-level contribution from $\Uc_0$ at leading order; here, this is the three-point function, which diagrammatically reads:
\begin{align}
\!\expval{\phi_{\qv_1}\phi_{\qv_2}\phi_{\qv_3}}^\mathrm{c} &=
\begin{tikzpicture}[baseline={(0,-0.1)}]
    \begin{feynman}
        \vertex (a) at (0,0);
        \vertex[dot, minimum size=3.5pt, label={[label distance=.2em]right:$\hat{u}_0^{(1)}$}] (b) at (0.5,0) {};
        \vertex (c) at (0.933,0.5);
        \vertex (d) at (0.933,-0.5);
        \diagram* {
            (a) -- (b),
            (b) -- (c),
            (b) -- (d)
        };
    \end{feynman}
\end{tikzpicture}
+ \order{\alpha^3}\nonumber\\
&= \!-\frac{6\alpha\noise^2\dbar_{\qv_1+\qv_2+\qv_3}}{\prod_{i=1}^3\Lc_{\qv_i}} \hu^{(1)}_0(\qv_1,\qv_2,\qv_3)\!+\!\order{\alpha^3},\label{eq:lambda_threept2}
\end{align}
with $\hu_0^{(1)}$ given in \eqref{eq:uAMB}. This corresponds to \eqref{eq:mpt_implicit} for $m=2$, $n=1$. The three-point function in AMA has the same form with $\hu_0^{(1)}$ from \eqref{eq:uAMA}; both are shown in Fig.~\fref[b]{mfig:AMAB} for $d=1$.

On the other hand, from the vertices of \eqref{eq:AMB_vertices}, we see that the Gaussian two-point function is only corrected at order $\alpha^2$; diagrammatically, up to this order, we write:
\begin{align}\label{eq:AMB_2pt}
\expval{\phi_{\qv}\phi_{\qv'}}^\mathrm{c} &=
\begin{tikzpicture}[baseline={(0,-0.1)}]
    \begin{feynman}
        \vertex (a) at (0,0);
        \vertex (b) at (1,0);
        \diagram* {
            (a) -- (b)
        };
    \end{feynman}
\end{tikzpicture}
+
\begin{tikzpicture}[baseline={(0,-0.1)}]
    \begin{feynman}
        \vertex (a) at (-0.3,0);
        \vertex[dot, minimum size=3.5pt, label={[label distance=-.2em]below left:$\hat{u}_0^{(1)}$}] (b) at (0.3,0) {};
        \vertex[dot, minimum size=3.5pt, label={[label distance=-.2em]below right:$\hat{u}_0^{(1)}$}] (c) at (0.7,0) {};
        \vertex (d) at (1.3,0);
        \diagram* {
            (a) -- (b) -- [half left] (c) -- (d),
            (c) -- [half left] (b)
        };
    \end{feynman}
\end{tikzpicture}
+
\begin{tikzpicture}[baseline={(0,-0.1)}]
    \begin{feynman}
        \vertex[empty dot, minimum size=3.5pt, label=below:$\hat{u}_{0}^{(2)}$] (m) at (0, 0) {};
        \vertex (a) at (-0.5,0);
        \vertex (b) at ( 0.5,0);
        \diagram* {
            (a) -- (m),
            (m) -- (b),
            m -- [loop, min distance=0.5cm] m,
        };
    \end{feynman}
\end{tikzpicture}
+
\begin{tikzpicture}[baseline={(0,-0.1)}]
    \begin{feynman}
        \vertex[red, empty dot, minimum size=3.5pt, label=below:$\hat{u}_{1}^{(2)}$] (m) at (0, 0) {};
        \vertex (a) at (-0.5,0);
        \vertex (b) at ( 0.5,0);
        \diagram* {
            (a) -- (m),
            (m) -- (b),
        };
    \end{feynman}
\end{tikzpicture}\nonumber\\
&=\expval{\phi_{\qv}\phi_{\qv'}}^\mathrm{c}_0+\expval{\phi_{\qv}\phi_{\qv'}}^\mathrm{c}_{\alpha^2}+\order{\alpha^4}\,,
\end{align}
where $\expval{\phi_{\qv}\phi_{\qv'}}^\mathrm{c}_0$ is the Gaussian two-point function, given by
\begin{equation}
\expval{\phi_{\qv}\phi_{\qv'}}^\mathrm{c}_0=
\begin{tikzpicture}[baseline={(0,-0.1)}]
    \begin{feynman}
        \vertex (a) at (0,0);
        \vertex (b) at (1,0);
        \diagram* {
            (a) -- (b)
        };
    \end{feynman}
\end{tikzpicture}
= \frac{\noise}{\Lc_\qv}\dbar_{\qv+\qv'}\,,
\end{equation}
and $\expval{\phi_{\qv}\phi_{\qv'}}^\mathrm{c}_{\alpha^2}$ is the correction from the nonlinearity.

Applying our Feynman rules to the diagrams of \eqref{eq:AMB_2pt}, we find the expressions:
\begin{subequations}
    \begin{align}
    \begin{tikzpicture}[baseline={(0,-0.1)}]
    \begin{feynman}
        \vertex (a) at (-0.3,0);
        \vertex[dot, minimum size=3.5pt, label={[label distance=-.2em]below left:$\hat{u}_0^{(1)}$}] (b) at (0.3,0) {};
        \vertex[dot, minimum size=3.5pt, label={[label distance=-.2em]below right:$\hat{u}_0^{(1)}$}] (c) at (0.7,0) {};
        \vertex (d) at (1.3,0);
        \diagram* {
            (a) -- (b) -- [half left] (c) -- (d),
            (c) -- [half left] (b)
        };
    \end{feynman}
    \end{tikzpicture}
    &= \frac{18\alpha^2\noise^2\dbar_{\qv+\qv'}}{\Lc_\qv^2}\int \ddbar{\tilde \qv} \frac{[\hat u^{(1)}_0(\qv,\tilde\qv,-\qv-\tilde\qv)]^2}{\Lc_{\tilde\qv+\qv}\Lc_{\tilde\qv}}\,,\label{eq:twopt_saturn}\\
    \begin{tikzpicture}[baseline={(0,-0.1)}]
    \begin{feynman}
        \vertex[empty dot, minimum size=3.5pt, label=below:$\hat{u}_{0}^{(2)}$] (m) at (0, 0) {};
        \vertex (a) at (-0.5,0);
        \vertex (b) at ( 0.5,0);
        \diagram* {
            (a) -- (m),
            (m) -- (b),
            m -- [loop, min distance=0.5cm] m,
        };
    \end{feynman}
    \end{tikzpicture}
    &= -\frac{12\alpha^2\noise^2\dbar_{\qv+\qv'}}{\Lc_\qv^2}\int \ddq \frac{\hat u^{(2)}_0(\qv,-\qv,\tilde\qv,-\tilde\qv)}{\Lc_{\tilde\qv}}\,,\label{eq:twopt_black}\\
    \begin{tikzpicture}[baseline={(0,-0.1)}]
    \begin{feynman}
        \vertex[red, empty dot, minimum size=3.5pt, label=below:$\hat{u}_{1}^{(2)}$] (m) at (0, 0) {};
        \vertex (a) at (-0.5,0);
        \vertex (b) at ( 0.5,0);
        \diagram* {
            (a) -- (m),
            (m) -- (b),
        };
    \end{feynman}
    \end{tikzpicture}
    &= -\frac{2\alpha^2\noise^2\dbar_{\qv+\qv'}}{\Lc_\qv^2}\hu_1^{(2)}(\qv,-\qv)\,.\label{eq:twopt_red}
    \end{align}
\end{subequations}
Using \eqref{eq:srn} and \eqref{eq:s_to_u}, we find:
\begin{widetext}
    \begin{align}\label{eq:AMB_2pt_corr}
    \expval{\phi_{\qv}\phi_{\qv'}}^\mathrm{c}_{\alpha^2} = \frac{6\alpha^2\noise^2\dbar_{\qv+\qv'}}{\Mc_\qv\Lc_\qv^2}\,\int\ddbar{\tilde\qv}\, \frac{-\Lc_{\tilde\qv+\qv}\hat s_{0}^{(2)}(\qv,-\qv,\tilde\qv,-\tilde\qv)+3\Mc_\qv\Lc_\qv[\hat u^{(1)}_0(\qv,\tilde\qv,-\qv-\tilde\qv)]^2}{\Lc_\qv\Lc_{\tilde\qv}\Lc_{\tilde\qv+\qv}}\,.
    \end{align}
    Equation \eqref{eq:sn} also results in:
    \begin{align}
    \hat s_0^{(2)}(\qv,-\qv,\tilde\qv,-\tilde\qv) ={}& -\left[\hat K(\qv,\tilde\qv)\hu_0^{(1)}(-\qv,-\tilde\qv,\qv+\tilde\qv)+\hat K(\qv,-\tilde\qv)\hu_0^{(1)}(-\qv,\tilde\qv,\qv-\tilde\qv)\right]\nonumber\\
    &{}-3\Biggl[\Mc_{\qv+\tilde\qv}\hu_0^{(1)}(-\qv,-\tilde\qv;\qv+\tilde\qv)^2+\Mc_{\qv-\tilde\qv}\hu_0^{(1)}(-\qv,\tilde\qv;\qv-\tilde\qv)^2\Biggr]\,,
    \end{align}
    which in \eqref{eq:AMB_2pt_corr}, using the variable transformation $\tilde \qv \to -\tilde \qv$ where appropriate, leads to
    \begin{align}
    \expval{\phi_{\qv}\phi_{\qv'}}^\mathrm{c}_{\alpha^2} = \frac{6\alpha^2\noise^2\dbar_{\qv+\qv'}}{\Mc_\qv\Lc_\qv^2}\,\int\ddbar{\tilde\qv}\, \frac{2\Lc_{\tilde\qv+\qv}\hat K(\qv,\tilde\qv)+6(\Mc_{\qv+\tilde\qv}\Lc_{\tilde\qv+\qv}+2\Mc_\qv\Lc_\qv)\hat u^{(1)}_0(\qv,-\qv,\tilde\qv,-\tilde\qv)}{\Lc_\qv\Lc_{\tilde\qv}\Lc_{\tilde\qv+\qv}}\hat u^{(1)}_0(\qv,-\qv,\tilde\qv,-\tilde\qv)\,.
    \end{align}
    Using \eqref{meq:u} and the variable transformation $\tilde\qv\to -\tilde\qv-\qv$ where needed, we finally obtain:
    \begin{align}\label{eq:AMB_2pt_result}
    \expval{\phi_{\qv}\phi_{\qv'}}^\mathrm{c}_{\alpha^2} =&-\frac{6\alpha^2\noise^2\dbar_{\qv+\qv'}}{\Mc_\qv\Lc_\qv^2}\,\int\ddbar{\tilde\qv}\,\frac{\hat K(\tilde\qv,-\qv-\tilde\qv)\hat u^{(1)}_0(\qv,-\qv,\tilde\qv,-\tilde\qv)}{\Lc_{\tilde\qv}\Lc_{\tilde\qv+\qv}}\,.
    \end{align}
\end{widetext}
Due to the non-analyticity of $\hu_0^{(1)}$, the integral in \eqref{eq:AMB_2pt_result} is likewise non-analytic as $\qv\to\vb{0}$. We find that its non-analytic part scales as $\expval{\phi_{\qv}\phi_{\qv'}}^\mathrm{c}_{\alpha^2}\sim q^{2(g-2)+d}$ (up to logarithmic factors), resulting in a real-space algebraic decay $\expval{\phi(\vb{0})\phi(\yv)}^\mathrm{c}_{\alpha^2}\sim |y|^{-2[g-2+d]}$ (and $\sim |y|^{-8}$ for AMB+ in $d=1$).

\subsection{Scaling of general \texorpdfstring{$\N$}{N}-point functions}
\label{sec:correlator_scaling}
For general $m$, we can make the following argument on the scaling of the two-point function. Its calculation to order $\order{\alpha^2}$ will contain a vertex
\begin{equation}\label{eq:twopt_twoptvertex}
\begin{tikzpicture}[baseline={(0,-0.1)}]
\begin{feynman}
\vertex[red, empty dot, label=below:$\hat{u}_{m-1}^{(2)}$] (m) at (0, 0) {};
    \vertex (a) at (-0.5,0);
    \vertex (b) at ( 0.5,0);
    \diagram* {
        (a) -- (m),
        (m) -- (b),
    };
\end{feynman}
\end{tikzpicture}\,,
\end{equation}
since $\m(m-1,2)=2$. Due to \eqref{eq:urn_scale_SM}, it scales as $\expval{\phi_{\qv}\phi_{\qv'}}^\mathrm{c}_{\alpha^2}\sim \alpha^2\noise^{\nonlin}q^{2(g-2)+d(m-1)}$. In real space, this results in $\expval{\phi(\vb{0})\phi(\yv)}^\mathrm{c}_{\alpha^2}\sim \alpha^2\noise^{\nonlin}|y|^{-[md+2(g-2)]}$, as given in \eqref{meq:2pt}.

Note that all other vertices contributing at order $\order{\alpha^2}$ besides \eqref{eq:twopt_twoptvertex} have the same scaling in $q$, as is the case in \eqref{eq:AMB_2pt}: considering lower values of $r$ removes factors $q^{dr}$ in $\hu_r^{(2)}$ in \eqref{eq:urn_scale_SM} but reintroduces the same factor in the additional loop integrals required to contract the new pairs of legs (cf.~\eqref{eq:twopt_black}); likewise, considering $n=1$ instead of $n=2$ means a factor $q^{g-2}$ is lost from \eqref{eq:urn_scale_SM} which is regained by having to consider two such vertices (cf.~\eqref{eq:twopt_saturn}).

For even $\nonlin$, the two-point function does not have any contribution linear in $\alpha$. However, there are such contributions if $\nonlin$ is odd. While each of these is long-ranged, they sum up to an overall term at order $\order{\alpha}$ that is short-ranged for the conserved class we consider. Although this can be proven via the diagrammatic expansion, it is significantly easier to show that this fact holds via the dynamical calculation discussed in Sec.~\ref{sec:dyncalc}, as we do below \eqref{eq:twopt_shortranged}. 

General $\N$-point functions with $\N>2$ instead scale as follows. Firstly, for $\N = (\nonlin-1)n + 2$ for some $n\in \mathbb{N}$, the leading-order contribution is the tree-level term \eqref{eq:mpt_implicit}, which with \eqref{eq:urn_scale_SM} scales as $q^{n(g-2)}$. Any further contribution (say, from the diagrams at $n+1$) has either loop integrals that give additional factors of $q$ or comes from larger $r$, again giving additional factors due to \eqref{eq:urn_scale_SM}.

For correlators with $\N \neq (\nonlin-1)n + 2$, we find that vertices with this number of fields first appear at $n_* = \lceil(\N - 2)/(\nonlin - 1)\rceil$ and $r_* = [(m-1)n_*- (\N-2)]/2$, thus leading to an $\N$-point function scaling as $q^{n_*(g-2)+dr}$ (the loop contributions at the same order $\alpha^{n_*}$ have the same scaling in $q$). Note that for odd $\nonlin$, the $\phi\to -\phi$ symmetry of the theory is preserved and odd $\N$-point functions vanish exactly.

\subsection{Direct calculation of \texorpdfstring{$\N$}{N}-point functions from their dynamics}
\label{sec:dyncalc}
\subsubsection{General \texorpdfstring{$\N$}{N}-point functions}
We now show an alternative way to derive expressions for the stationary $\N$-point functions, by obtaining them directly from Eq.~\eqref{meq:dynamics} without knowledge of the stationary measure, order by order in $\alpha$. From that equation, the Itô dynamics for the $\N$-point function of Eq.~\eqref{eq:npt} can be extracted, yielding:
\begin{align}
\partial_t F_\N ={}& \N\left[(\alpha\Kc_{\qv_\N}-\Mc_{\qv_\N}\Lc_{\qv_\N}\phi_{\qv_\N})F_{\N-1}\brq{N-1}\right]^\symm\nonumber\\
&{}+ \noise\,\N(\N-1)\left[\Mc_{\qv_\N}\dbar_{\qv_{\N-1}+\qv_{\N}}F_{\N-2}\brq{N-2}\right]^\symm\nonumber\\
&{}+ \N\left[\sqrt{2\noise\Mc_{\qv_\N}}\eta_{\qv_\N}F_{\N-1}\brq{N-1}\right]^\symm,
\end{align}
where the noise $\eta$ is zero-mean Gaussian, obeying $\expval{\eta_\qv\eta_{\qv'}}=\dbar_{\qv+\qv'}\dbar(t-t')$.

Taking the average, and requiring stationarity, we find (for $\sum_{i=1}^\N\Mc_{\qv_i}\Lc_{\qv_i} \neq 0$):
\begin{align}
\expval{F_\N} ={}& \frac{\noise \N(\N-1)\Bigl[\expval{F_{\N-2}\brq{\N-2}}\dbar_{\qv_{\N-1}+\qv_\N}\Mc_{\qv_\N}\Bigr]^\symm}{\sum_{i=1}^\N\Mc_{\qv_i}\Lc_{\qv_i}}\nonumber\\
&{}+\alpha\frac{\N\expval{F_{\N-1}\brq{\N-1}\Kc_{\qv_\N}}^\symm}{\sum_{i=1}^\N\Mc_{\qv_i}\Lc_{\qv_i}}\,.\label{eq:lpt}
\end{align}
The nonlinearity couples the $\N$-point function to higher correlators by means of the second term. Thus, Eq.~\eqref{eq:lpt} constitutes an infinite hierarchy of coupled $\N$-point functions; this hierarchy can be solved perturbatively in $\alpha$.

\subsubsection{AMB+ correlators}
We now show that this procedure gives the same results as the diagrammatic calculation  of Sec.~\ref{sec:loop} for AMB+. We start from the three-point function of \eqref{eq:lambda_threept2}. Equation~\eqref{eq:lpt} gives:
\begin{align}
\expval{F_3\brq{3}} ={}& \frac{6\noise \left[\expval{\phi_{\qv_1}}\dbar_{\qv_2+\qv_3}\Mc_{\qv_3}\right]^\symm}{\sum_{i=1}^3\Mc_{\qv_i}\Lc_{\qv_i}}\nonumber\\
&{}+\frac{3\alpha\expval{F_{2}\brq{2}\Kc_{\qv_3}}^\symm}{\sum_{i=1}^3\Mc_{\qv_i}\Lc_{\qv_i}}\,.
\end{align}
Note that the one-point function $\expval{\phi_{\qv}}$ vanishes whenever $\qv\neq \vb{0}$ due to translational invariance. When all $\qv_i\neq\vb{0}$, we thus have $\expval{F_3\brq{3}}=\expval{F_3\brq{3}}^\mathrm{c}$ and, using \eqref{meq:K},
\begin{align}\label{eq:3pt_dyn}
\expval{F_3\brq{3}}^\mathrm{c} ={}& \frac{3\alpha\expval{\phi_{\qv_1}\phi_{\qv_2}\Kc_{\qv_3}}^\symm}{\sum_{i=1}^3\Mc_{\qv_i}\Lc_{\qv_i}}\nonumber\\
={}& \frac{3\alpha}{\sum_{i=1}^3\Mc_{\qv_i}\Lc_{\qv_i}}{}\cdot{}\nonumber\\
\cdot\int\pddqt{2}&\expval{\phi_{\qv_1}\phi_{\qv_2}\phi_{\tilde\qv_1}\phi_{\tilde\qv_2}\dbar_{\qv_3-\tilde\qv_1-\tilde\qv_2}}^\symm\hat K(\tilde \qv_1,\tilde \qv_2)\,.
\end{align}
To lowest order in $\alpha$, we can approximate the four-point function in \eqref{eq:3pt_dyn} with that of the Gaussian theory, which can be evaluated using Wick's theorem to give:
\begin{align}\label{eq:3pt}
\expval{F_3\brq{3}}^\mathrm{c}
&= \frac{6\alpha\noise^2\dbar_{\qv_1+\qv_2+\qv_3}}{\sum_{i=1}^3\Mc_{\qv_i}\Lc_{\qv_i}}{}\cdot{}\nonumber\\
\cdot\int&\pddqt{2}\Biggl(\frac{\dbar_{\qv_1+\tilde\qv_2}\dbar_{\tilde\qv_1+\qv_2}}{\Lc_{\qv_1}\Lc_{\tilde\qv_1}}\Biggr)^\symm\hat K(\tilde \qv_1,\tilde \qv_2)+\order{\alpha^2}\nonumber\\
&= \frac{6\alpha\noise^2\dbar_{\qv_1+\qv_2+\qv_3}}{\sum_{i=1}^3\Mc_{\qv_i}\Lc_{\qv_i}}\frac{[\hat K(\qv_1,\qv_2)\Lc_{\qv_3}]^\symm}{\Lc_{\qv_1}\Lc_{\qv_2}\Lc_{\qv_3}}+\order{\alpha^2}\nonumber\\
&=-\frac{6\alpha\noise^2\dbar_{\qv_1+\qv_2+\qv_3}}{\prod_{i=1}^3\Lc_{\qv_i}}\hu_0^{(1)}(\qv_1,\qv_2,\qv_3)+\order{\alpha^2}\,,
\end{align}
where in the last line we have used \eqref{meq:u} to retrieve the result of Eq.~\eqref{meq:mpt}.

We now turn to the two-point function of AMB+ and show that we recover the result of \eqref{eq:AMB_2pt_result} for $\nonlin=2$. Equation~\eqref{eq:lpt} gives (for $\qv\neq\vb{0}$):
\begin{align}
\expval{\phi_{\qv}\phi_{\qv'}}^\mathrm{c}
={}&{}\frac{\noise}{\Lc_{\qv}}\dbar_{\qv+\qv'}+\alpha\frac{2\expval{\phi_{\qv}\Kc_{\qv'}}^\symm}{\Mc_{\qv}\Lc_{\qv}+\Mc_{\qv'}\Lc_{\qv'}}\,.\label{eq:twopt}
\end{align}
The first term is the equilibrium contribution from the unperturbed Gaussian dynamics, while the second term is the correction from the nonlinearity, which using \eqref{meq:K} and \eqref{eq:3pt} yields to order $\order{\alpha^2}$:
\begin{align}
\expval{\phi_\qv\phi_{\qv'}}_{\alpha^2}^\mathrm{c}={}& \frac{\alpha\dbar_{\qv+\qv'}}{\Mc_{\qv}\Lc_{\qv}}\int\pddqt{2}\expval{\phi_{\qv}\phi_{\tilde\qv_1}\phi_{\tilde\qv_2}}\hat K(\tilde \qv_1,\tilde \qv_2)\nonumber\\
={}& -\frac{6\alpha^2\noise^2\dbar_{\qv+\qv'}}{\Mc_{\qv}\Lc_{\qv}}{}\cdot{}\nonumber\\
\cdot\int\ddbar{\tilde\qv}&\frac{\hu_0^{(1)}(\qv_1,\tilde\qv_1,-\qv_1-\tilde\qv_1)\hat K(\tilde \qv_1,-\qv_1-\tilde \qv_1)}{\Lc_{\qv}\Lc_{\tilde\qv}\Lc_{\tilde\qv+\qv}},
\end{align}
a result which coincides with \eqref{eq:AMB_2pt_result}.

\subsubsection{Linear correction to the two-point function}
For odd $m$, the two-point function has a contribution that is linear in $\alpha$. We now show that this contribution, for models in the conservative class we consider, is short-ranged. Indeed, for this case \eqref{eq:twopt} with \eqref{meq:K} gives:
\begin{align}
\expval{\phi_\qv\phi_{\qv'}}={}& \frac{\noise}{\Lc_{\qv}}\dbar_{\qv+\qv'}\nonumber\\
&{}+ \frac{\alpha\dbar_{\qv+\qv'}}{\Mc_{\qv}\Lc_{\qv}}\int\pddqt{m}\expval{\phi_{\qv}\prod_{j=1}^m\phi_{\tilde\qv_j}}\hat K(\tilde\Qv_m)\,.
\end{align}
To order $\order{\alpha}$, the $(m+1)$-function in the integral can be evaluated with respect to the unperturbed Gaussian measure; using Wick's theorem, this gives
\begin{align}\label{eq:twopt_shortranged}
\expval{\phi_\qv\phi_{\qv'}}={}& \frac{\noise}{\Lc_{\qv}}\dbar_{\qv+\qv'}\nonumber\\
&{} + \frac{\alpha\dbar_{\qv+\qv'}}{\Mc_{\qv}\Lc_{\qv}}\frac{m(m-1)!}{2^{(m-1)/2}((m-1)/2)!}\cdot{}\nonumber\\
\cdot\int\hspace{-.2em}\pddqt{(m-1)/2}&\hat K(-\qv,\tilde\qv_1,-\tilde\qv_1,\ldots,\tilde\qv_{(m-1)/2},-\tilde\qv_{(m-1)/2})\,.
\end{align}
For conserved dynamics, the $\Mc_\qv = q^2$ in the denominator is compensated by a factor from the integral: Indeed, for a conservative nonlinearity, $\hat K(\Qv_m)$ has a factor $\Qvs{m}$ (coming from the divergence in $\Kc=-\div\Jv$), which here is equal to $-\qv$. This factor can be taken out of the integral, which becomes a vectorial quantity; due to isotropy, the only vector it can be proportional to is $\qv$, resulting in the sought factor $q^2$. Thus, overall \eqref{eq:twopt_shortranged} is analytic and two-point correlations are short-ranged at $\order{\alpha}$.

\section{Multiple fields}
We now specify how the considerations from the single-field case generalize to theories with $\mathfrak{N}$ fields. Consider the dynamics
\begin{equation}\label{eq:dynamics_multi}
 \partial_t\phi^{a}_\qv = -\sum_b\hat B^{ab}_\qv\phi^{b}_\qv + \alpha\Kc^{a}_{\qv} + \sqrt{2\noise\Mc^a_\qv}\eta^a_\qv\,,
\end{equation}
where $a, b=1,\ldots,\mathfrak{N}$ are field indices. No summation convention is used in this Section. The noises $\eta^a_\qv$ are zero-mean Gaussian, satisfying $\expval{\eta^a_\qv(t)\eta^b_{\qv'}(t')}=\delta^{ab}\dbar_{\qv+\qv'}\delta(t-t')$.

We aim at obtaining the stationary measure of \eqref{eq:dynamics_multi} along the lines of Sec.~\ref{sec:perturbative}. To this goal, we require the quasipotential of the unperturbed dynamics, $\Uc_0^{(0)}$, and its instanton $\Phiv^{(0)}(t)$. The former is easily found: for $\alpha=0$, Eq.~\eqref{eq:dynamics_multi} describes an Ornstein-Uhlenbeck process, for which the stationary measure satisfies ${P_\infty[\phiv]\propto\exp(-\Uc_\mathrm{G}/\noise)}$ with
\begin{equation}\label{eq:multi_stationary}
 \Uc_0^{(0)}[\phiv] = \Uc_\mathrm{G}[\phiv] = \frac{1}{2}\int \ddq \sum_{a,b}\phi^a_{-\qv}\Oc^{ab}_{\qv}\phi^b_{\qv}\,.
\end{equation}
Here, for every $\qv$, $\Omatrix_\qv$ is a symmetric matrix obeying the Lyapunov equation~\cite{risken-1996}:
\begin{equation}\label{eq:lyapunov}
\frac{1}{2}\left[\vu B_\qv\vdot\Omatrix_\qv^{-1} + \Omatrix_\qv^{-1}\vdot\vu B_\qv^\dagger\right]^{ab} = \Mc_\qv^a\delta^{ab}\,,
\end{equation}
Furthermore, the instanton of the unperturbed dynamics solves~\cite{freidlin-2012}:
\begin{align}
\partial_t\Phiv^{(0)}_\qv = \vu{A}_\qv\vdot\Phiv^{(0)}_\qv\,,\quad\Phiv^{(0)}_\qv(t=0) = 
\phi^a_\qv\,,\label{eq:instanton_multi}
\end{align}
with $\Phi^a_\qv(t=-\infty) = 0$ for non-conserved fields and $\Phi^a_\qv(t=-\infty) = \phi^a_\qv\delta_{\qv,\vb{0}}$ for conserved fields, and
\begin{equation}\label{eq:hatA}
    \hat A^{ab}_\qv \coloneqq 2\Mc^a_\qv{\Oc}^{ab}_\qv - \hat{B}^{ab}_\qv\,.
\end{equation}
Equation \eqref{eq:instanton_multi} is solved by
\begin{equation}
 \Phiv^{(0)}_\qv(t) = \exp(\vu{A}_\qv t)\vdot\phiv_\qv\,,\vspace{.5em}\label{eq:instanton_multi_sol}
\end{equation}
where $\exp$ is the matrix exponential.
The perturbative corrections to the stationary measure in the expansion of \eqref{meq:WKB} are then given by
\begin{equation}\label{eq:Uc_S_multi}
\Uc_r^{(n)}[\phiv] = \int_{-\infty}^0\dd{t}\Sc_r^{(n)}[\Phiv^{(0)}(t)]\,,
\end{equation}
with
\begin{align}
&\Sc_0^{(n)}[\phiv] ={}\nonumber\\
&{}\quad-\int\ddq\sum_a\left[\mu^{a,(n-1)}_{-\qv}\Kc^a_\qv + \sum_{k=1}^{n-1}{\mu^{a,(n-k)}_{-\qv}}\Mc^a_\qv{\mu^{a,(k)}_{\qv}}\right],
\end{align}
where $\mu^{a,(n)}_\qv = (2\pi)^d\fdv{\Uc_0^{(n)}}{\phi^a_{-\qv}}$, and, for $r>0$,
\begin{align}
\Sc_r^{(n)}[\phiv] ={}&\int\dd{\qv}\sum_a
\Biggl[\Mc^a_\qv\fdv{\mu^{a,(n)}_{r-1,\qv}}{\phi^a_{\qv}}+\fdv{\Kc^a_\qv}{\phi^a_\qv}\delta_{n,1}\delta_{r,1}\Biggr]\nonumber\\
&{}-\int\ddq\sum_a\Biggl[\mu^{a,(n-1)}_{r,-\qv}\Kc^a_\qv\nonumber\\
&{}\quad+2\sum_{k=1}^{n-1}{\mu^{a,(n-k)}_{0,-\qv}}\Mc^a_\qv{\mu^{a,(k)}_{r,\qv}}\nonumber\\
&{}\quad+\sum_{k=0}^{n}\sum_{s=1}^{r-1}{\mu^{a,(n-k)}_{r-s,-\qv}}\Mc^a_\qv{\mu^{a,(k)}_{s,\qv}}\Biggr]\,.
\end{align}
The lowest-order correction to \eqref{eq:multi_stationary} thus reads
\begin{align}\label{eq:U1_multi}
\Uc_0^{(1)} ={}-\int_0^\infty\dd{t}\int\ddq\sum_{a,b}\Oc_{\qv}^{ab}\Phi^{b,(0)}_\qv(t)\Kc^a_\qv[\Phiv^{(0)}(t)]\,.
\end{align}
Using \eqref{eq:instanton_multi_sol}, the time integral in \eqref{eq:U1_multi} (and, generally, \eqref{eq:Uc_S_multi}) can be solved explicitly, as we show below for the case $\mathfrak{N} = 2$. Note that also for multiple fields, the number $\mrn$ of fields in $\Uc_r^{(n)}$ is still given by \eqref{eq:mrn} (although different combinations of fields are now allowed).

\subsection{Two fields}
Here, we show how to apply the above to the case ${\mathfrak{N}=2}$, showing that NRCH and the minimal Toner--Tu model of the main text have a nonlocal stationary measure. First, let us note that if $\vu{A}_\qv$ has distinct eigenvalues $\lambda_a(\qv)$, the matrix exponential in \eqref{eq:instanton_multi_sol} can be expressed by means of Sylvester's formula as
\begin{subequations}\label{eq:sylvester}
    \begin{align}
    \exp(\vu{A}_\qv t) &= e^{\lambda_+(\qv)t}\vu{A}^+_\qv-e^{\lambda_-(\qv)t}\vu{A}_\qv^-\,,\\
    \vu{A}^\pm_\qv&=\frac{\vu{A}_\qv-\lambda_\mp(\qv)\id}{\lambda_+(\qv)-\lambda_-(\qv)}\,.\label{eq:Ahatpm}
    \end{align}
\end{subequations}
\subsubsection{Non-reciprocal Cahn-Hilliard model}
We first treat the dynamics of Eq.~\eqref{meq:NRCH}, and show that their stationary measure is nonlocal when a nonlinearity is present. In Fourier space, this model takes the form of \eqref{eq:dynamics_multi} with $\phiv = (\NRCHone, \NRCHtwo)$, $\Mc^a_\qv=\Mc_\qv$, and $\hat B_\qv^{ab} = \Mc_\qv\Lc_\qv^{ab}$, where $\Mc_\qv=q^2$. Detailed balance is observed iff $\Lc_\qv^{ab} =\Lc_\qv^{ba}$ (reciprocal case), in which case $\Oc^{ab}_\qv = \Lc^{ab}_\qv$. On the other hand, if $\Lc^{ab}_\qv$ is not symmetric, the Gaussian model is already out of equilibrium, with $\Oc^{ab}_\qv \neq \Lc^{ab}_\qv$.

If we assume $\Kc_\NRCHone=\order{\nabla^g\NRCHone^{m-s}\NRCHtwo^s}$, we find the following scalings. Firstly, we have $\Omatrix_\qv\sim \order{q^0}$ from \eqref{eq:lyapunov}. Furthermore, $\hat A^{ab}_\qv\sim \Mc_\qv(2\Oc^{ab}_\qv-\Lc^{ab}_\qv)\sim \order{q^2}$ and thus $\lambda_\pm(\qv) \sim \order{q^2}$, while $\vu{A}^\pm_\qv\sim \order{q^0}$ in~\eqref{eq:Ahatpm}. Thus, in \eqref{eq:U1_multi}, the only factors of $q$ come from $\Kc$ and from the time integration over the instanton of \eqref{eq:instanton_multi_sol}. The former gives a factor $q^g$, while the time integration results with \eqref{eq:sylvester} in an inverse sum over the $\lambda_\pm(\qv_i)\sim q_i^2$ (as we will see explicitly in \eqref{eq:instanton_integral}). The resulting numerator and denominator don't cancel out in general, so that we obtain a non-analytic $\hat u_0^{(1)}\sim q^{g-2}$ (corresponding to $\beta=g-2$ and $\G_0^{(1)}\sim |y|^{-[md+g-2]}$), and thus a nonlocal stationary measure.

To make these arguments more concrete, we now show explicit results for the simplest nonreciprocal case with $\Lc^{ab}_\qv = \Lc_\qv\delta^{ab} + \Delta\varepsilon^{ab}$, where $\varepsilon$ is the Levi--Civita symbol and $\Lc_\qv = 1 + \xi^2 q^2$. Solving the Lyapunov equation \eqref{eq:lyapunov} for this case leads to $\Oc^{ab}_\qv = \Lc_\qv\delta^{ab}$, so that \eqref{eq:multi_stationary} gives
\begin{equation}\label{eq:NRCH_stationary}
 \Uc_\mathrm{G}[\NRCHone,\NRCHtwo] = \frac{1}{2}\int \ddq \Lc_{\qv}\left(|\NRCHone_\qv|^2+|\NRCHtwo_\qv|^2\right).
\end{equation}
Thus, in this simple case the Gaussian stationary measure only depends on the symmetric (reciprocal) part of $\Lc^{ab}_\qv$, and the fields are uncoupled. The instanton is also particularly simple in this case. We find from \eqref{eq:hatA} that $\hat A^{ab}_\qv = \Mc_\qv(\Lc_\qv\delta^{ab} - \Delta\varepsilon^{ab})$, which has eigenvalues $\lambda_\pm(\qv) = \Mc_\qv(\Lc_\qv\pm i\Delta)$. We thus obtain from \eqref{eq:instanton_multi_sol}:
\begin{align}\label{eq:instanton_NRCH}
    &\Phiv^{(0)}_{\qv}(t) = {}\nonumber\\
    &\,\frac{1}{2}e^{\Mc_\qv\Lc_\qv t}\Bigg[e^{i\Mc_\qv\Delta t}\mqty(1&i\\-i&1)+e^{-i\Mc_\qv\Delta t}\mqty(1&-i\\i&1)\Bigg]\vdot\mqty(\NRCHone_\qv\\\NRCHtwo_\qv)\,,
\end{align}
We consider the NRCH nonlinearity $\Kc_\density = \laplacian \NRCHone^{\nonlin}$ (the standard choice is $m=3$). In Fourier space, the nonlinearity has the form
\begin{equation}
    \Kc^\NRCHone_\qv = -\Mc_\qv\int\pddq{\nonlin}\NRCHone_{\qv_i}\dbar_{\qv-\Qvs{\nonlin}}\,,
\end{equation}
and we find from \eqref{eq:U1_multi}:
\begin{equation}
 \Uc_0^{(1)}=\int\pddq{{\nonlin+1}}[\Mc_{\qv_1}\Lc_{\qv_1}]^\symm\dbar_{\Qvs{\nonlin+1}}\int_0^\infty\!\dd{t}\,\prod_{j=1}^{\nonlin+1}\Phi^{\NRCHone,(0)}_{\qv_j}(t)\,.
\end{equation}
From the expression \eqref{eq:instanton_NRCH} for the instanton,
we see that for $\Delta \neq 0$, the fields $\NRCHone$ and $\NRCHtwo$ become coupled in the stationary measure; the nonlinearity in $\NRCHone$ even affects the coupling of $\NRCHtwo$ to itself.
We thus find the structure:
\begin{equation}\label{eq:NRCH_U_struc}
 \Uc_0^{(1)}=\!\sum_{l=0}^{\nonlin+1}\!\int\pddq{{\nonlin+1}}\prod_{j=1}^l\NRCHtwo_{\qv_j}\prod_{\mathclap{k={l+1}}}^{{\nonlin+1}}\NRCHone_{\qv_k}\hu_{0,l}^{(1)}\brq{m+1}\,\dbar_{\Qvs{\nonlin+1}}\,,
\end{equation}
where, using that
\begin{align}\label{eq:instanton_integral}
\int_0^\infty\!\dd{t}\,\prod_{j=1}^{\nonlin+1}\Phi^{\NRCHone,(0)}_{\qv_j}(t) &= \frac{1}{2^{\nonlin+1}}\sum_{\mathclap{\{s_j = \pm\}}}\frac{\prod_{j=1}^{\nonlin+1}(\NRCHone_{\qv_j}+is_j\NRCHtwo_{\qv_j})}{\sum_{j=1}^{\nonlin+1}\lambda_{s_j}}\,,
\end{align}
the kernels are given by
\begin{align}\label{eq:NRCH_generic_l}
&\hu_{0,l}^{(1)}\brq{m+1}=\binom{{\nonlin+1}}{l}\frac{i^l}{(\nonlin+1)2^{\nonlin+1}}\cdot{}\nonumber\\
&\cdot\sum_{\{s_j = \pm\}}\frac{\left[\sum_{j=1}^{\nonlin+1}\Mc_{\qv_j}\Lc_{\qv_j}\right]\prod_{j=1}^l s_j}{\left[\sum_{j=1}^{\nonlin+1}\Mc_{\qv_j}\Lc_{\qv_j}\right]+i\Delta\left[\sum_{j=1}^{\nonlin+1}s_j\Mc_{\qv_j}\right]}\,.
\end{align}
Note that these kernels are real (as follows from the complex-conjugate structure of the square bracket in \eqref{eq:instanton_NRCH}).
For $\Delta \neq 0$, all $\hu_{0,l}^{(1)}$ are non-analytic at the origin, with $\hu_{0,l}^{(1)} \sim q^0$, resulting in algebraically decaying real-space couplings $\G_{l}^{(1)}(\Yv_{\nonlin}) \sim |y|^{-\nonlin d}$, which have $\beta=0$ in agreement with the general treatment above. Note that as in the single-field case, here again the conservation law is essential, since for $\Mc_\qv \sim \order{q^0}$ the couplings would be analytic.

Because both fields are conserved, the scalings with $q$ are in fact the same as for the single-field case throughout the hierarchy of $\Uc_r^{n}$. This implies that the $\hu_r^{n}$ follow the scaling of Eq.~\eqref{eq:urn_scale_SM} also in NRCH.

We finally note that the correlators can be obtained from the $\Uc_r^{(n)}$ by generalizing the single-field Feynman rules of Sec.~\ref{sec:feynman}; while we don't present this explicitly, this means that the arguments on the scaling of the correlators from the single-field case straightforwardly extend to NRCH, so that the results of Sec.~\ref{sec:correlator_scaling} still hold for this model.

\subsubsection{Minimal Toner--Tu}
We now turn to the dynamics of Eqs.~\eqref{meq:rhop}, showing that here as well adding a nonlinearity leads to a nonlocal stationary measure. In the notation of \eqref{eq:dynamics_multi}, we have $\Mc^\density_q = 0$, $\Mc^\pol_q = 1$, and
\begin{equation}\label{eq:B_TT}
\vu{B}_q = \mqty(0 & iq\vel\\
iq \prhoc & \Lc^\pol_q)\,.
\end{equation}
Solving \eqref{eq:lyapunov} yields
\begin{equation}
\Omatrix_q = \Lc^\pol_q\mqty(\omega_\density & 0\\0 & \omega_\pol),
\end{equation}
where $\omega_\density = -\prhoc/\vel$ and $\omega_\pol = 1$.
Calculating $\vu{A}_q$ from \eqref{eq:hatA} reveals that:
\begin{align}
\tr \vu{A}_q &= \Lc^\pol_q\sim \order{q^0}\,,\\
\det \vu{A}_q &= q^2\vel\prhoc\sim \order{q^2}\,,
\end{align}
so that from
\begin{equation}
\lambda_\pm(q) = \frac{\tr \vu{A}_q}{2}\left[1\mp \sqrt{1-4\frac{\det \vu{A}_q}{(\tr \vu{A}_q)^2}}\right]
\end{equation}
we obtain $\lambda_- \sim\order{q^0}$ and $\lambda_+ \sim\order{q^2}$. The latter scaling reflects the presence of one conserved field in the dynamics.

We now consider nonlinearities $\Kc_{\hat a} = \order{\partial^g_x \density^{\nonlin-s}\pol^{s}}$, with $\hat a \in \{\density, \pol\}$ indicating which equation $\Kc$ is added to. In Fourier space, we find:
\begin{equation}
    \Kc^{\hat a}_q[\phi] ={} \int\prod_{i=1}^{s}[\ddbar{q_i}\,\pol_{q_i}]\prod_{j=s+1}^{\nonlin}[\ddbar{q_j}\,\density_{q_i}]\hat K(\Qv_\nonlin)\,\dbar_{q-q^\Sigma_\nonlin}\,,
\end{equation}
where $\hat K$ is a polynomial of order $g$. Let us introduce unit vectors $\vu{e}_i$ in field space corresponding to the fields appearing in $\Kc_{\hat a}$ (\emph{i.e.}, $\vu{e}_{1,\ldots,s} = (0,1)$ and $\vu{e}_{s+1,\ldots,\nonlin} = (1,0)$), and $\vu{e}_{\nonlin+1}$ corresponding to the field $\hat a$.
We then find from \eqref{eq:U1_multi}:
\begin{align}
&\Uc_0^{(1)} =\nonumber\\
&-\!\int_0^\infty\!\!\!\dd{t}\!\int\prod_{i=1}^{\nonlin+1}\left[\ddbar{q_i}\,\vu{e}_{a_i}\vdot \Phiv^{(0)}_{q_i}(t)\right]\Lc^\pol_{q_{\nonlin+1}}\omega_{\hat a}\hat K\brq{\nonlin}\,\dbar_{q^\Sigma_{\nonlin+1}}\,.
\end{align}
Now, we are interested in the nonlocal part $\Uc_\mathrm{0,NL}^{(1)}$ of the quasipotential. We thus only consider the contribution from $\lambda_+(q)\sim q^2$ in Eq.~\eqref{eq:sylvester} (since contributions involving $\lambda_-$ will be local, cf.~\eqref{eq:instanton_integral}), obtaining:
\begin{align}
&\Uc_\mathrm{0,NL}^{(1)}=\nonumber\\
&-\int\,\prod_{i=1}^{\mathclap{\nonlin+1}}\left[\ddbar{q_i}\,\vu{e}_{a_i}\vdot\vu{A}^+_{q_i}\vdot \phiv_{q_i}\right]\frac{\hat K\brq{\nonlin}\,\Lc^\pol_{q_0}\omega_{a_0}\,\dbar_{q^\Sigma_{\nonlin+1}}}{\sum_{i=1}^{\nonlin+1}\lambda_+(q_i)}\,.\label{eq:UNL_rhop}
\end{align}
The denominator vanishes at the origin; this again generically results in a non-analyticity, yielding a nonlocal $\Uc_0^{(1)}$. Because the matrix $\vu{A}^+_q$ mixes the fields, we find a structure akin to \eqref{eq:NRCH_U_struc}:
\begin{equation}
 \Uc_\mathrm{0,NL}^{(1)}=\sum_{l=0}^{\nonlin+1}\!\int\prod_{i=1}^{\nonlin+1}\ddbar{q_i}\,\prod_{j=1}^l\pol_{q_j}\prod_{\mathclap{k={l+1}}}^{{\mathclap{\nonlin+1}}}\density_{q_k}\hu_{0,l}^{(1)}\brq{m+1}\,\dbar_{q^\Sigma_{\nonlin+1}}\,.
\end{equation}
To obtain the scaling of each $\hu_{0,l}^{(1)}$, we note that:
\begin{align}
\vu{A}^+_{k} \sim \mqty(\order{q^0} & \order{q}\\\order{q} & \order{q^2})\,.
\end{align}
Thus, the scaling of the contraction in the round bracket of Eq.~\eqref{eq:UNL_rhop} depends on which fields appear in $\Kc_{\hat a}$, where the nonlinearity is added to (\emph{i.e.}, $\hat a$), and which fields are being coupled: for each polarization $\pol$ in either position, an extra factor of $q$ is generated. Together with the $q^{g-2}$ from the fraction, we obtain $\hu_{0,l}^{(1)} \sim q^{l+s+g-2}$ for a density nonlinearity $\Kc_\density$, and $\hu_{0,l}^{(1)} \sim q^{l+s+g-1}$ for a polarization nonlinearity $\Kc_\pol$. This results in real-space scaling $\G_l^{(1)}(\Yv_{\nonlin})\sim |y|^{-[m+l+s+g-2]}$ and $\G_l^{(1)}(\Yv_{\nonlin})\sim |y|^{-[m+l+s+g-1]}$, respectively.

As examples for $\Kc_{\hat a}$, consider the advective polarization nonlinearity $\Kc_\pol=- \partial_x p^2$, which has $\hu_{0,l}^{(1)} \sim q^{l+2}$, the advective density nonlinearity $\Kc_\density=-\partial_x(\density\pol)$, which has $\hu_{0,l}^{(1)} \sim q^{l}$, and the nonlinearity $\Kc_\pol=-\pol^{\nonlin}$, leading to $\hu_{0,l}^{(1)}\sim q^{\nonlin+l-1}$. In the case of the advective density nonlinearity, choosing $l=0$, we find $\beta=0$, so that the Landau--Ginzburg expansion for $\density$ fails at bulk level.

We finally note that adding conservative noise to the density dynamics, $\Mc^\density_q=q^2$, and a diffusive deterministic term, so that instead of \eqref{eq:B_TT} we have:
\begin{equation}
\vu{B}_q = \mqty(q^2\Lc^\density_q & iq\vel\\
iq \prhoc & \Lc^\pol_q)\,,
\end{equation}
with $\Lc^\density_q\sim \order{q^0}$, does not affect any of the scalings given here.

\bibliography{../../references}